\documentclass{emulateapj}
\def\arcsec{$^{\prime\prime}$}
\newcommand\degree{{^\circ}}
\newcommand\surfb{$\mathrm{mag}/\square$\arcsec}
\newcommand\kms{\rm{~km~s}^{-1}}
\newcommand\Gyr{\rm{~Gyr}}
\newcommand\msun{\rm{M}_\odot}

\shorttitle{Structure of Thick Disks}
\shortauthors{Yoachim \& Dalcanton}

\begin{document}

\title{Structural Parameters of Thin and Thick 
Disks in Edge-On Disk Galaxies}

\author{Peter Yoachim\altaffilmark{1} 
  \&  Julianne J. Dalcanton\altaffilmark{2,3}}
\affil{Department of Astronomy, University of Washington, Box 351580,
Seattle WA, 98195}

\altaffiltext{1}{e-mail address: \email{yoachim@astro.washington.edu}}
\altaffiltext{2}{Alfred P. Sloan Foundation Fellow}
\altaffiltext{3}{e-mail address: \email{jd@astro.washington.edu}}

\begin{abstract}
We analyze the global structure of 34 late-type, edge-on, undisturbed,
disk galaxies spanning a wide range of mass.  We measure structural
parameters for the galaxies using two-dimensional least-squares
fitting to our $R$-band photometry.  The fits require both a thick and
a thin disk to adequately fit the data.  The thick disks have larger
scale heights and longer scale lengths than the embedded thin disks,
by factors of $\sim$2 and $\sim$1.25, respectively.  The observed
structural parameters agree well with the properties of thick and thin
disks derived from star counts in the Milky Way and from resolved
stellar populations in nearby galaxies. We find that massive galaxies'
luminosities are dominated by the thin disk.  However, in low mass
galaxies (V$_{\rm{c}} \lesssim 120$ km s$^{-1}$), thick disk stars
contribute nearly half of the luminosity and dominate the stellar
mass.  Thus, although low mass dwarf galaxies appear blue, the
majority of their stars are probably quite old.

Our data are most easily explained by a formation scenario where the
thick disk is assembled through direct accretion of stellar material
from merging satellites while the thin disk is formed from accreted
gas.  The baryonic fraction in the thin disk therefore constrains the
gas-richness of the merging pre-galactic fragments.  If we include the
mass in HI as part of the thin disk, the thick disk contains
$\lesssim\!10$\% of the baryons in high mass galaxies, and
$\sim\!25-30$\% of the baryons in low-mass galaxies.  Our data
therefore indicate that the fragments were quite gas rich at the time
of merging ($f_{gas}=75-90$\%).  However, because low mass galaxies
have a smaller fraction of baryons in their thin disks, the
pre-galactic fragments from which they assembled must have been
systematically more gas poor.  We believe this trend results from
increased outflow due to supernova-driven winds in the lower mass
pre-galactic fragments.  We estimate that $\sim\!60$\% of the total
baryonic mass in these systems was lost due to outflows.  Pushing the
episode of significant winds to early times allows the
mass-metallicity relationship for disks to be established early,
before the main disk is assembled, and obviates the difficulty in
driving winds from diffuse disks with low star formation efficiencies.
We discuss other implications of this scenario for solving the G-dwarf
problem, for predicting abundance trends in thick disks, and for
removing discrepancies between semi-analytic galaxy formation models
and the observed colors of low mass galaxies.
\end{abstract}

\keywords{galaxies: formation --- galaxies: structure}

\section{Introduction}\label{intro}

The structure of galactic disks provides strong constraints on their
formation and evolution.  Spiral galaxies have long been recognized to
contain several distinct populations of stars (e.g., disks, bulges and
halos), each with distinct chemical and kinematic properties that
capture unique epochs in the formation of the galaxy.  Observations of
the Milky Way and a wide range of other galaxies have revealed the
need for yet another component, namely, a thick stellar disk.
Originally detected as an excess of light at high galactic latitudes
in deep surface photometry of early-type galaxies \citep{Burstein79,
Tsikoudi79}, a thick disk was later revealed in the Milky Way using
star counts \citep{Gilmore83}.

The properties of the Milky Way's thick disk have revealed many
differences from the thin disk.  Structurally, the Milky Way's thick
disk has a significantly larger scale height than the thin disk, as
its name implies \citep[for reviews see][and references
therein]{Reid93, Buser99, Norris99}.  It also may have a somewhat
longer scale length \citep{Robin96, Ojha01, Chen01, Larsen03}.  Thick
disk stars are older and more metal-poor than stars in the thin disk
\citep[e.g.][]{Reid93, Chiba00}.  They are also significantly enhanced
in $\alpha$-elements, compared to thin disk stars of comparable iron
abundance \citep{Fuhrmann98,Prochaska00, Taut01, Bensby03, Feltzing03,
Mishenina04, Brewer04, Bensby05}.  Kinematically, Milky Way thick disk
stars have both larger velocity dispersions and slower net rotation
than stars in the thin disk \citep{Nissen95, Chiba00, Gilmore02,
Soubiran03, Parker04}.

For many years, however, it remained unclear whether the thick disk
was a truly distinct component of the Milky Way, or whether it was
only an older, metal-poor extension of the thin disk, as might be
created by steady vertical heating over the lifetime of the Galaxy
\citep[e.g., ][]{Dove93}.  Over the past five years, conclusive
evidence that the thick disk is indeed distinct from the thin disk has
come from a series of detailed chemical abundance studies.  Stars with
thick disk kinematics show significant alpha-enhancement compared to
thin disk stars with identical iron abundances, thus forming a
separate parallel sequence in a plot of [$\alpha$/H] vs [Fe/H]
\citep[see the recent review by][]{Feltzing04}.  Studies of resolved
stars in nearby galaxies also find a thick disk of old red giant
branch stars whose lack of metallicity gradient cannot be explained by
steady vertical heating \citep{Seth05b, Mould05}.

Three general classes of formation mechanisms have been proposed to
explain the properties of the Milky Way thick disk.  In the first, a
previously thin disk is dynamically heated to form a thick disk, after
which a new thin disk forms \citep{Quinn93,Velazquez99, Robin96,
Chen01}.  In the second, the thick disk forms directly from gas at a
large scale height, possibly during a largely monolithic
proto-galactic collapse \citep{ELS,Gilmore86, Norris91, Burkert92,
Kroupa02, Fuhrmann04, Brook04}.  In the third, the thick disk forms
from a series of minor-merger events which directly deposit stars at
large scale heights \citep{Statler88,Abadi203}.  Recent cosmological
simulations have suggested a more complicated origin.  Disk galaxy
simulations by \citet{Abadi203} find a thick disk which is composed
primarily of tidal debris from disrupted satellites while comparable
simulations by \citet{Brook04} find that thick disk stars form during
a period of chaotic mergers of gas-rich building blocks.  Recent
kinematic measurements favor scenarios where mergers play a
significant role in thick disk formation \citep{Gilmore02, Yoachim05}.

While all of the above scenarios are viable explanations for the
origin of the Milky Way, the structural parameters of thin and thick
disk components in a wide range of galaxies can help distinguish among
these formation scenarios.  Unfortunately, the measurements required
to characterize thick disks are difficult to make outside the Milky
Way.  The Milky Way thick disk provides less than 10\% of the local
stellar density \citep{Buser99}, and this faintness hampers detailed
study of comparable extragalactic thick disks.  To date, thick disk
structural properties have been measured only in a small number of
galaxies \citep[see Table~2 below]{Seth05b,Pohlen04,
vanDokkum94, Morrison97, Neeser02, Abe99, Wu02,deGrijs96,deGrijs97b}.
These studies analyze galaxies in the edge-on orientation, which
allows clear delineation between regions where thin and thick disk
stars dominate the flux. The edge-on orientation also allows line of
sight integrations of faint stellar populations to reach detectable
levels.

In this paper, we analyze a large sample of edge-on galaxies and
decompose them into thick and thin disk components. Analysis of $B$,
$R$, \& $K_s$ photometry and color maps has previously revealed that
these galaxies are surrounded by a flattened faint red envelope, with
properties very similar to the Milky Way thick disk
\citep{Dalcanton02}.  We now use a full 2-dimensional fitting
procedure capable of simultaneously fitting the thick and thin disk
light distributions to derive their full structural parameters.

\subsection{Galaxy Sample}

The sample used in this paper was drawn from the optical and infrared
imaging found in \citet[hereafter Paper I]{Dalcanton00}.  Briefly, our
sample of edge-on bulgeless galaxies was initially selected from the
Flat Galaxy Catalog (FGC) of \citet{Karachentsev93}, a catalog of 4455
edge-on galaxies with axial ratios greater than 7 and major axis
lengths grater than $0.6^\prime$.  The color maps and initial
detections of the thick disks in 47 galaxies were presented in
\citet[hereafter Paper II]{Dalcanton02}.  

Not all galaxies from Paper I have been included in the analysis
presented here.  We have excluded several of the more massive galaxies
with sizable bulge components that could not be adequately masked or
modeled.  We likewise eliminated several low mass galaxies with bright
central star clusters for similar reasons.  We have also removed any
galaxies that have either significant warps or visible spiral arms
(i.e., that were not viewed perfectly edge-on), as these systems are
poorly modeled by our fitting procedure.  Finally, we eliminated
galaxies whose surface brightness profiles would be severely affected
by atmospheric seeing.  A full list of the 15 excluded galaxies are
listed in Table~\ref{rej_table}, leaving a sample of 34 galaxies
suitable for decomposing into thick and thin components.  When
possible, we have used distances listed in \citet{Kara00} derived from
a local flow model.  Otherwise we use the galaxy's redshift corrected
for the motion of the Local Group \citep{Yahil77}, assuming a Hubble
Constant of $H_0=70$ km s$^{-1}$ Mpc$^{-1}$.

\begin{deluxetable}{ l l }
\tablewidth{0pt}
\tablecaption{{Galaxies rejected from the 2-disk $R$-band fitting} 
\label{rej_table}}
\tablehead{\colhead{FGC(E)} & {Reason for rejection}}
\startdata
51 & Large central knot \\
84 & Spiral arms visible\\
143 & Bright central star-forming region\\
442 & Spiral arms visible \\
256 & Thin disk below seeing limit, fits did not converge\\
1863 & Large warp, bright foreground stars \\
1945 & Fits did not converge, possibly spiral arms \\
1971 & Polar ring galaxy\\
2217 & Large bulge component\\
2367 & Spiral arms visible\\
2264 & Fits did not converge, scattered light problem\\
2292 & Thin disk below seeing limit, bright foreground stars\\
E1440 & Asymmetric disk \\
E1447 & No velocity data\\
E1619 & Large bulge component
\enddata
\end{deluxetable}

\section{2D Fitting }

\subsection{Galaxy Models \label{models}}

The distinctive vertical color gradients identified in Paper II
suggest that the stellar population above the galaxies' midplanes is
different from the one within it. We assume this change is due to the
existence of two distinct stellar populations 
analogous to the MW's thick and thin disks.  Our 2-dimensional fitting
procedure attempts to decouple these two populations to measure their
scale heights, scale lengths, and luminosities.

We model the surface brightness of each disk component as a radially
exponential disk.  We adopt the luminosity density $\mathcal{L}$ of
each disk component to be
\begin{equation}
 \mathcal{L}(R,z)=\mathcal{L}_0 \mathrm{e}^{-R/h_R} f(z)
\end{equation}
where $(R,z)$ are cylindrical coordinates, $\mathcal{L}_0$ is the central
luminosity density, $h_R$ is the radial scale length, and $f(z)$ is
a function describing the vertical distribution of stars.  

Throughout, we adopt a generalized vertical distribution 
\begin{eqnarray}
f(z)=\mathrm{sech}^{2/N}(Nz/z_0) \label{z3}
\end{eqnarray}
where $z_0$ is the vertical scale height and $N$ is a parameter
controlling the shape of the profile near the midplane.  For
appropriate choices of $N$, this equation can reproduce many popular
choices for the vertical distribution of star light.  With $N=1$
Equation~\ref{z3} becomes the expected form for a self-gravitating
isothermal sheet \citep{Spitzer42,vdk81b,vdk81a, vdk82}.  When
$N\rightarrow\infty$, Equation~\ref{z3} reduces to
$f(z)\propto\mathrm{e}^{-z/h_z}$, where $h_z=z_0/2$.  Previous fits to
the vertical distribution suggest that an intermediate value of $N=2$
is a better model of galaxy disks \citep{vdk88}, as expected for the
superposition of several populations with a range of vertical velocity
dispersions \citep{deGrijs96}.  However, different values of $N$ only
produce differences near the galaxy midplane, and all share
exponentially declining profiles at large radii.

When fitting a thick plus thin disk model, we preferentially use $N=1$
for both components because of its physical motivation.  We note that
since our main goal is not to model galaxies near their midplane where
these functions have their largest differences, our results are not
particularly sensitive to the choice of model.  To permit comparisons
to previous work, we also derive single disk fits to our sample using
Equation~\ref{z3} without a fixed $N$, allowing the shape of the
vertical profile to vary to best fit the data.

To translate the adopted luminosity density into the observed surface
brightness distribution, we assume that the disks are viewed perfectly
edge-on.  Other authors have demonstrated that slight deviations from
$i=90\degree$ have minimal impact on the derived structural parameters
\citep[e.g.,][]{vdk81a,deGrijs97a}.  We also assume that scale heights
are independent of projected radius for late-type galaxies, as found
by \citet{vdk81a,Bizyaev02,deGrijs97b,Shaw90}.  With the above
assumptions, the model edge-on disk surface brightness is given by
\begin{equation} \label{sech2}
\Sigma(R, z) = \Sigma_{0,0}
(R/h_{R})K_{1}(R/h_{R}) f(z)
\end{equation}
where $K_{1}$ is a modified Bessel function of the first order, $
\Sigma_{0,0}$ is the edge-on peak surface brightness ($\Sigma_{0,0}=2
h_{R} \mathcal{L}_0$), and $R$ is now the projected radius along the
major axis.  The face-on surface brightness of such a disk is
$\Sigma(R)=\Sigma_{0}e^{-R/h}$ with $\Sigma_{0}=2 z_0 \mathcal{L}_0$.
Throughout, we convert our edge-on peak surface brightnesses to
magnitudes using $\mu(0,0)=m_{zp}-2.5\mathrm{log} (\Sigma_{0,0})$
where $m_{zp}$ is the photometric zero point from Paper I.  The
face-on central surface brightness can then be calculated as $\mu_0=
\mu(0,0)- 2.5\mathrm{log} (z_0/h_R)$.  The conversion between the edge-on
and face-on orientation assumes that disks are optically thin at any
orientation, an assumption that is obviously not true for massive
galaxies with dust lanes.  However, we correct for this effect later
in \S\ref{dust_effects}.  We do not model any possible disk
truncation, as this is a small effect seen only in the region $R > 3
h_R$ \citep{vdk81a,Kregel04,Pohlen00}.

Our sample of galaxies was initially selected to be ``pure disk''
systems, and thus there are very few galaxies which possess a
prominent bulge component.  We therefore do not attempt to decompose a
bulge component from the surface brightness distribution and simply
reject galaxies with significant bulges from the sample
(Table~\ref{rej_table}).

We have tested if the profiles described by Equation~\ref{z3} could be
significantly affected by seeing.  We convolved model images with a
two-dimensional circular Gaussian kernel to simulate the atmosphere's
effect.  We found that this step, in general, was unnecessary.
Unconvolved fits differed from convolved fits only for the most distant
galaxies.  Several of these galaxies have been eliminated from the
sample, as listed in Table~\ref{rej_table}.

%#####################SUBSECTION FITTING METHOD###################
\subsection{Fitting Method}\label{sec_fitting}

We use Levenberg-Marquardt least squares fitting of the galaxy images
to find the best parameters for the models described in
\S\ref{models}.  Before fitting, the images of the galaxies are
sky-subtracted and foreground stars and background galaxies are
generously masked (see Paper II).  The images are cropped at $R \sim 4
h_R$ to speed computation time. Our tests have shown that the fits are
insensitive to the exact cropping, causing variations in individual
parameters of only a few percent.  The cropping also reduces the
chance that our fits could be biased by warps or flaring of the disks
at large radii.

Following the technique of \citet{Kregel02}, we weight each pixel by
the inverse of the model surface brightness distribution at that
pixel. By using the model rather than the data to determine the
weighting, we eliminate the bias of overweighting positive noise
spikes.  This weighting scheme places large amounts of weight on the
lowest signal-to-noise pixels, ensuring that regions of low surface
brightness (i.e., where a faint thick disk could be detected) receive
adequate weighting.  To prevent the fit from being overwhelmed by
regions with low signal-to-noise, we set the weight to zero beyond the
1-$\sigma$ noise contour, defined as where the model falls below the
standard deviation of the background.  Due to the low signal-to-noise
ratio in the $K_s$ band data, these images were clipped at the
1/2-$\sigma$ level to ensure that an adequate number of pixels were
included in the fit.  Each fit was iterated four times to ensure
convergence of the model parameters and weighting scheme.  Fits were
performed using pixel coordinates and counts, then converted to
arcseconds and magnitudes using the calibrations in Paper
I.%\citet{Dalcanton00}.

It is common practice when fitting models to edge-on galaxies to crop
out regions near the midplane of the disk \citep[e.e.,][]{Kregel02,
Jong96, Bizyaev02}.  Cropping avoids the hard-to-model effects of dust
lanes, bulges, and star forming regions.  The color maps of our
galaxies imply that galaxies rotating at speeds less than 120 km
s$^{-1}$ do not contain concentrated central dust lanes
\citep{Dalcanton04}.  For more massive galaxies, our weighting
procedure ensures that any midplane structure receives a minimal
amount of weighting when calculating the goodness-of-fit $\chi^2$.  We
chose to fit models both with and without the midplane cropping to
quantify the systematic uncertainties introduced by midplane
structure.

We begin by fitting single disk models to all three $B$, $R$, and, $K_s$
images, holding the galaxy position and rotation fixed.  We then fit
2-component models to the images, allowing the offsets and rotation to
vary, but constraining all components to have the same center and
orientation.  For this second step, we use only the $R$-band images
due to their high signal-to-noise.  Ideally, we would perform the
2-disk decomposition in the $K_s$ band which best represents the
smooth stellar distribution and is least effected by dust.  However,
due to the bright infrared sky, the NIR images are lower
signal-to-noise and cannot reach to faint regions where a thick disk
would dominate.  The $R$-band therefore represents the best compromise
between reaching faint regions of the galaxies while minimizing the
effects of dust extinction and bright star forming regions.

When fitting a single disk with only three free parameters, our
procedure converges to the same $\chi^2$ minima given any reasonable
initial guesses.  However, for multiple component models, which have
up to 10 free parameters, we find that fits often converge to local
minima rather than to the global minimum.  To ensure we find the
global minimum when fitting multiple components, we fit each galaxy
using up to 50 unique initial parameter guesses, following
\citet{Wu02}.  The initial parameters for each galaxy model were
randomly varied up to $\pm 50\%$ to ensure we cast a large net in
parameter space.

The formal parameter uncertainties that result from our fits are not
meaningful because we used a weighting scheme that is not based on the
actual pixel uncertainties.  Even if we did minimize $\chi^2$ using
formal pixel errors, our returned uncertainties would be much too low.
The $\chi^2$ formalism requires residuals to be Gaussian, which is
rarely the case when fitting nearby galaxies.  The situation is
comparable to trying to model Mount Rainier as a cone--you can do it,
but the residuals will be dominated by real physical structures and
not random Gaussian measurement errors.  In the case of spiral
galaxies, real substructure exists in the form of spiral arms, dust
lanes, regions of active star formation, warps, flares, HII regions,
etc.  As an alternative assessment of the systematic errors which are
likely to dominate our uncertainties, we fit a series of models using
a variety of different weighting and masking schemes
(Table~4) and quote the median result for each
parameter.  We then adopt the full range of convergent models for each
parameter as a measure of the inherent systematic uncertainties.  The
resulting uncertainties are 2-100 times greater than our formal
$\chi^2$ uncertainties, confirming that systematic errors dominate our
uncertainties.

%########################SUBSECTION  TETS ON ARTIFICIAL IMAGES
\subsection{Tests on Artificial Images}\label{artif}

To assess the reliability of the 2-disk decompositions, we created a
set of 100 artificial galaxies.  We adopted the surface brightness
profile in Equation~\ref{z3} with an $N=1$ vertical distribution for
both a thick and thin component, and varied the structural parameters
of the disks within ranges similar to our sample galaxies (for the
thin disk: 20.7 \surfb 
$ < \mu(0,0) < 22.7$ \surfb, 2.5\arcsec$ < h_R
< 19.4$\arcsec, 0.6\arcsec$ < z_0 < 3.6$\arcsec; and the thick disk
had parameters in the range: 21.4 \surfb $< \mu(0,0) < 24.2$ \surfb,
1.5\arcsec$ < h_R < 51.2$\arcsec, 1.4\arcsec$ < z_0 < 20$\arcsec).
The model galaxies were convolved with a circular Gaussian with
1\arcsec ~FWHM, typical of the seeing for the observations.  We then
added read noise, sky noise, and Poisson noise to the simulated
galaxies, with amplitudes chosen to mimic our $R$-band data.  The
galaxy images were rotated up to 2 degrees and offset up to 2 pixels
(0.5\arcsec) from the image center.  We then fit the galaxies with the
two disk models, with and without seeing corrections.  We assume $N=1$
and use the same spread of initial parameter guesses as described in
\S\ref{sec_fitting}.

Of the 100 simulated galaxies, only three fits failed to converge.
90\% of the scale lengths are recovered to within $\pm$2\% of the
input value, with all of the results converging within $\pm10$\%.
90\% of the scale heights are recovered within $\pm$3\% of the input
value with all results within $\pm$12\%.  90\% of the central surface
brightnesses converge to within 0.09 mags of the correct result.  The
orientations were always correct to within 0.1 degrees, with a median
error less than 1\%.  All of the spatial offsets were within 1 pixel
of the correct position.  There were no systematic trends in the size
of the errors verses galaxy properties.  The high accuracy of these
fits indicates that we are not limited by pixel noise in our fits.
However, since our model was a perfect match to the input data, this
correspondence is not surprising.

We tested models that did not make a seeing correction convolution and
found the fits still returned scale lengths and heights that were
accurate to $< 0.2$\arcsec, as long as $z_0 > 1$\arcsec.  The majority
of our observed galaxies do have $z_0 > 1$\arcsec, and thus we do not
account for seeing in our fits.  This result is consistent with the
analysis of \citet{deGrijs97a} who find that for an exponential
vertical profile, convolution is unnecessary when the seeing FWHM $\le
0.6 h_z$.

In addition to testing our ability to recover the parameters of a
known model, we also tested our ability to correctly measure the
structural parameters when using an incorrect function for the
vertical light distribution.  Specifically, we fit an $N=2$ model
instead of the correct $N=1$ vertical profile to each disk.  These
fits returned results similar to the fits using the correct model, and
the resulting scale heights and lengths fell within $\sim 5\%$ of the
correct values.  This indicates that the galaxy sizes are constrained
primarily by light well away from the midplane.  The luminosities were
more divergent, however, due to the large differences between these
models at their midplanes.  The $N=2$ model was slightly biased
towards having over-luminous thick disks (with a few outliers as
well), but the majority (70\%) of fits were within a factor of 2 of
the correct $L_{thick}/L_{thin}$, despite being fit with the wrong
function.

Finally, we also tested our two-disk fitting code on artificial
galaxies that had no second thick disk component.  In these cases, the
fits always converged to extremely faint thick disks ($< 1\%$ of the
thin disk flux) and usually converged to either very large or very
small thick disk scale lengths, mimicking either a uniform sky
background or a small point source.  Overall, these results encourage
us to believe that if there are no thick disk components in our data
our fitted parameters will diverge to unphysical values.

%#####################SINGLE DISK FITS################
\section{Single Disk Fits}\label{sec_sd}

Before discussing the results of decomposing the galaxies into two
disk components (\S\ref{sec_2disks}), we discuss the results for
fitting single disks to the light distributions.  These fits are
useful simple descriptions of the galaxies, and the resulting
parameters can be directly compared to previous fits of edge-on and
face-on galaxies.

As discussed in \S\ref{sec_fitting}, we quantify our systematic
uncertainties using Equation~\ref{sech2} with an $N=1$ vertical
profile and a variety of weighting and masking techniques resulting in
five different fits for each galaxy image.  The five fits are: (1) the
full galaxy with inverse model weighting as described in \S
\ref{sec_fitting}; (2) the full galaxy with uniform weighting to more
heavily weight the high signal-to-noise regions; (3) inverse weighting
with the midplane region ($z=\pm z_0$) masked; (4) inverse weighting
with the outer region $R > 3 h_R$ masked, to eliminate regions where
our fit may be affected by unmodeled stellar truncation; and (5)
inverse weighting with the high latitude region $z > 2 z_0$ masked to
minimize the effect of thick disks.  The results of the fits are given
in Table~3 for all 3 band passes.  The columns show
the median edge-on peak surface brightnesses ($\mu(0,0)$), radial
scale lengths ($h_R$), and vertical scale parameters ($z_0$) along
with their uncertainties.  We emphasize again that these are not the
formal statistical uncertainties (which are deceptively small), but
instead are the full range of values to which the five different fits
converged.

\begin{figure}
\plotone{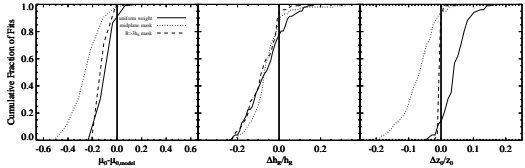}%\plotone{plots/single_cu.eps}
\caption{Cumulative distributions showing the systematic effects of
fitting our galaxies with varying models. All models are compared to a
fit using inverse weighting and no masked regions. \label{cumu_dist}
Fits for all three filters have been combined.  The solid curve shows
the model which used uniform pixel weighting, while the dotted curve
shows the midplane masked model, and the dashed curve shows the
$|R|<3h_R$ radially masked model.}
\end{figure}

In addition to quantifying our uncertainties, the five different fits
provide insight into how variations in fitting methods and weighting
schemes affect our results.  The systematic effects of the different
methods are plotted in Figure~\ref{cumu_dist}.  The most notable
features in Figure~\ref{cumu_dist} are the large systematic shifts in
the values of $z_0$ for the single disk fits.  Models that are
weighted to fit the midplane (e.g. the uniform weighting model),
return thinner disks while models that mask the midplane return larger
values of $z_0$.  This effect is present in all three filters, and is
exactly what one would expect if disk galaxies were dominated by thin
disks at their midplanes and by thick disks with larger scale-heights
in the fainter regions.  The radially cropped and midplane cropped
models result in fits that have fainter central regions and slightly
larger scale heights.  This is a strong indication that most of our
galaxies do not have dust lanes which need to be masked.  Cropping
regions at high $z$ has a minimal ($< 5 \%$ change) effect on the fit
parameters.

\begin{figure}
\plotone{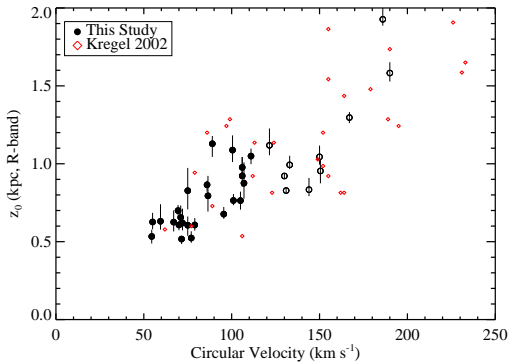}%\plotone{plots/single11.eps}
\caption{ Single disk scale heights for the $R$-band fits. Galaxies
with prominent dust lanes are plotted with with open circles.  For
comparison, we show the $I$-band scale heights from the edge-on
sample of \citet{Kregel02}, plotted as red diamonds. \label{single11}}
\end{figure}

\begin{figure}
\plottwo{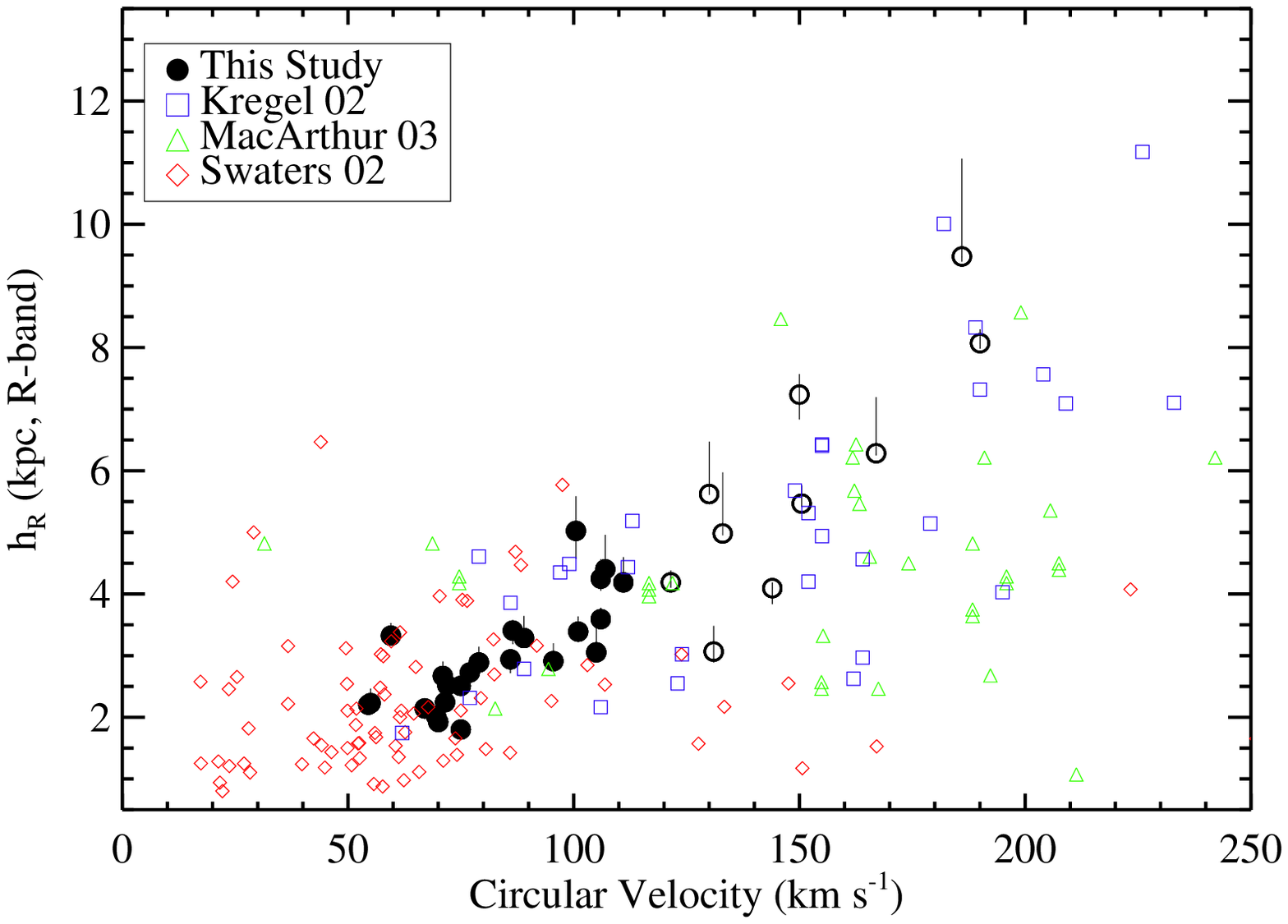}{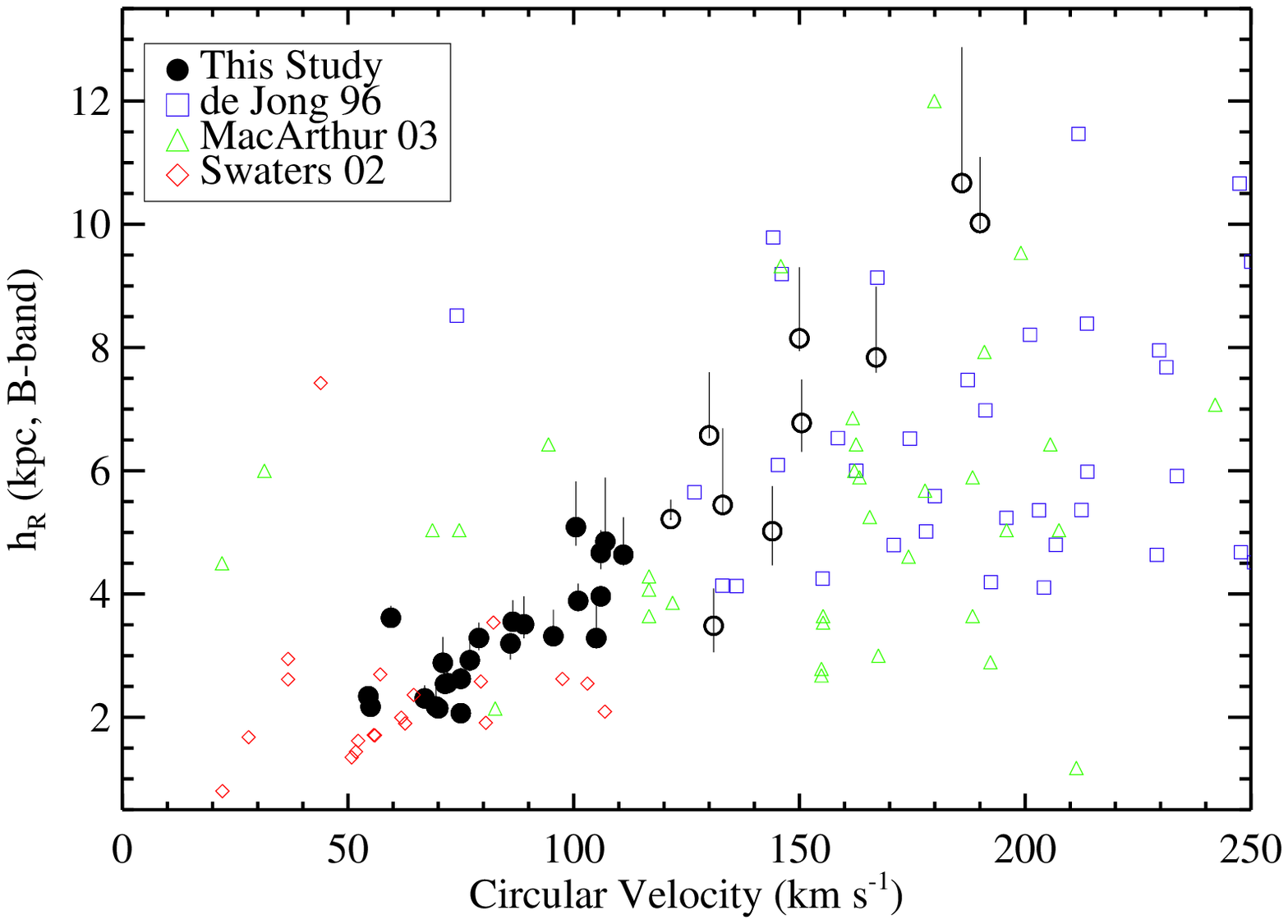}
%\plottwo{plots/hrsR.eps}{plots/hrsB.eps}
\caption{Single disk scale lengths for the $R$-band (left) and
$B$-band (right) fits. Galaxies with prominent dust lanes are plotted
with with open circles.  For comparison, we show other single disk
fits gathered from the literature.  The \citet{Jong96} and
\citet{MacArthur03} data are face-on or moderately inclined galaxy
samples with the scale lengths measured in the $R$ and $B$-bands
(plotted as blue squares and green triangles respectively).  The
\citet{Swaters02} sample consists of late-type spiral and irregular
galaxies with scale lengths measured in the $R$ and $B$-band (plotted
as red diamonds).  The \citet{Kregel02} data were measured from
edge-on galaxies in the $I$-band (plotted as blue squares).
\label{single1} }
\end{figure}

\begin{figure}
%\plotone{plots/single2.eps}
\plotone{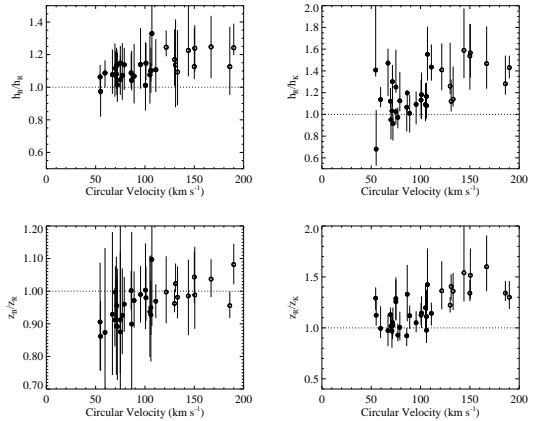}
\caption{ Comparison of scale lengths and heights for the single disk
fits in different bands. Open symbols are used for galaxies with
prominent dust lanes. \label{single2}}
\end{figure}

\begin{figure}
%\plotone{plots/single3.eps}
\plotone{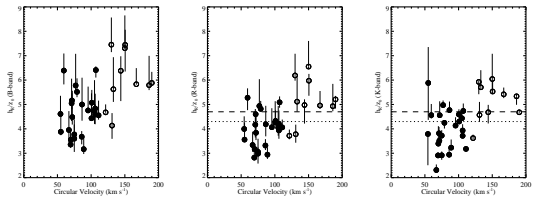}
\caption{Single disk fits showing the flatness ($h_R/z_0$) for each
band. Open symbols are used for galaxies with prominent dust lanes.
Dotted lines show the average flatness for a sample of 34 galaxies in
$I$-band presented in \citet{Kregel02}.  Dashed lines show the average
flatness measured measured from a sample of 153 galaxies from the
Revised Flat Galaxy Catalog imaged by 2MASS in the $K$-band and
presented in \citet{Bizyaev02} \label{single3}}
\end{figure}

\begin{figure}
%\plotone{plots/single41.eps}
\plotone{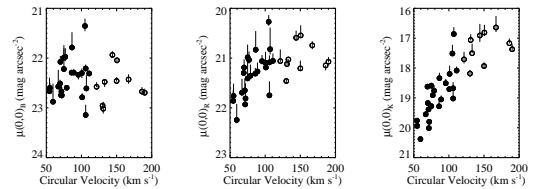}
\caption{ Edge-on peak surface brightnesses for the single disk
fits.  Open symbols are used for galaxies with prominent dust lanes.
\label{single4} Points have not been corrected for internal extinction.}
\end{figure}

The parameters for the single disk fits listed in Table~3 are plotted
in Figures~\ref{single11}, \ref{single1}, \ref{single2},
\ref{single3}, and \ref{single4} as a function of the galaxies'
circular velocity.  In Figure \ref{single11}, we see the expected
trends that more massive galaxies have larger scale heights.  We also
plot scale heights from the edge-on sample of \citet{Kregel02} and
find that both studies give consistent results for the scale height as
a function of galaxy circular velocity.  In Figure~\ref{single1} we
compare our single disk $R$-and $B$-band fitted scale lengths with the
edge-on sample of \citet{Kregel02} and the face-on measurements of
\citet{MacArthur03,Jong96}, and \citet{Swaters02}.  Overall, we find
that lower mass galaxies in the range $50 < V_c < 120$ have scale
lengths wholly consistent with measurements made in comparable face-on
systems.  However, the highest mass galaxies in our sample have scale
lengths that are slightly larger than the average found in previous
studies, although they are still within the full range of the
comparison data.  This offset is worse in the $B$-band than in the
$R$-band, and almost certainly reflects the presence of strong dust
lanes in the more massive systems.  The higher attenuation towards the
central regions of the galaxies will suppress the surface brightness
at small radii, leading to apparently larger scale lengths.  This
offset may also explain why studies of edge-on disks suggest that
disks truncate at only 3-4 $h_R$, whereas face-on studies see no
obvious signature of truncation at these radii \citep[ but see also
Pohlen et al. 2002\nocite{Pohlen02}]{Barton97, Weiner01}.

In Figure~\ref{single2} we compare the structural parameters derived
in different band passes.  We confirm that redder filters converge to
shorter scale lengths (Figure~\ref{single2}), a result of the strong
radial color gradients seen in both our sample and in face-on galaxies
\citep[e.g., ][]{Bell01, MacArthur04}.

We also find that for galaxies without dust lanes the $B$-band scale
heights are predominantly thinner than the $R$-band.  This offset is
consistent with the detection of strong vertical color gradients in
Paper II, where we found that the midplanes of late-type galaxies were
typically bluer than the light above the plane.  Somewhat
unexpectedly, our $K$-band scale heights are also significantly
thinner than the $R$-band.  We believe that this is due to three
effects.  First, the $K$-band data does not reach as deep as the other
filters, making it insensitive to the extended thick component.
Second, the thinner $K$-band scale height may indicate the presence of
dust which blocks light from the midplane in optical filters.
Finally, there is some indication from studies of resolved stars in
nearby galaxies that the $K$-band light is not completely dominated by
old red giant stars, but instead has a significant contribution from
young stars with small scale heights \citep{Seth05b}.

In Figure~\ref{single3} we plot the axial ratios of our sample
galaxies.  Overall, our axial ratios are consistent with the work of
\citet{Bizyaev02} who measure flatness parameters for 153 edge-on
galaxies imaged in the 2MASS survey and find values of $h_r/z_0$
ranging from $\sim2$ to $\sim10$.  Our results are also consistent
with the axial ratios from \citet{Kregel02}.  We see a slight trend
for more massive galaxies to be flatter than less massive galaxies.
Other studies have also suggested that low mass, low surface
brightness dwarf galaxies are thicker than regular spirals.  Estimates
of the intrinsic axial ratios of dwarf irregulars range from
$b/a\sim0.3$ \citep{Hodge66,Binggeli95} to $b/a\sim0.6$
\citep{Sung98,Staveley92}, all of which are rounder than typical
spirals \citep[e.g.][]{Kudrya94}.  We have discussed possible
explanations for this behavior in \citet{Dalcanton04}.

Figure~\ref{single4} shows the edge-on peak surface brightnesses for
the one-disk fits.  The peak surface brightness of the $B$-band data
is roughly constant, showing little trend with galaxy mass.  However,
because the FGC sample was initially selected from the POSS-II survey
plates, we would not expect the $B$-band surface brightnesses to be
below $\mu\sim 23$ \surfb.  On the brighter end, the Freeman law
\citep{Freeman70} suggests a maximum surface brightness for edge-on
disks.  Thus, the $B$-band peak surface brightnesses must be confined
to a limited range.  In contrast, we do see increasingly strong trends
of surface brightness with mass in the redder filters, and
particularly in $K_s$.  Because the selection criteria for the FGC
limited the range of $B$ surface brightness the observed trends in $R$
and $K$ are due to variations in galaxy color with mass.  As we will
discuss in \S\ref{dust_effects}, extinction from dust prevents us from
being able to reliably convert the edge-on brightnesses to comparable
face-on values.

%##########All the little single disk plots

%################Thick Disk Section#########
\section{Two Disk Fits}\label{sec_2disks}

\subsection{Need for a Second Component}\label{sec_need2}

\begin{figure}
%\plotone{plots/resids_splitm.eps}
\plotone{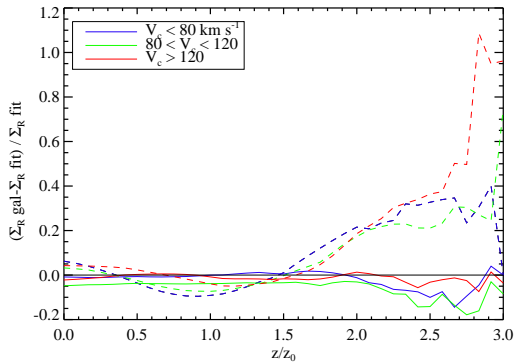}
\caption{ \emph{Left}: Residuals collapsed along the radial direction
and averaged over galaxies binned by mass.  The single disk models
(dashed lines) are a very poor fits, leaving large amounts of excess
flux at high latitudes.  The two disk models (solid lines) do a much
better job fitting the vertical light distribution at \emph{all}
latitudes and show only a small systematic trend to over-subtract at
high $z$.
\label{two_components}}
\end{figure}

The traditional signature of thick disks is the presence of excess
light at high latitudes after subtracting a single disk component.  To
demonstrate the expected excess, we subtract the single disk models
from the data and sum the residuals (inside the 1-$\sigma$ noise
contour) along the major axis.  The resulting residuals are plotted in
Figure~\ref{two_components}, and demonstrate that the single disk fits
from \S\ref{sec_sd} systematically leave excess flux at high latitudes
for all masses of galaxies. We also average the vertical profile
residuals across different galaxy mass ranges and find the two disk
model is superior to the single disk model in all cases.  For a single
disk model, we can slightly improve the fit at high $z$ by allowing
the index $N$ to vary.  However, on average, the absolute value of the
two-disk model residuals are smaller than the variable $N$ model at
\emph{every} height.  By collapsing along the radial direction, we are
assuming that any disk components have a nearly constant scale height
with radius, as has been found for late type disks in many studies
\citep{vdk81a,Bizyaev02,deGrijs97b}.

These tests show that (1) our galaxies are poorly fit at large $z$ by
the simple sech$^2$ model; (2) by leaving the index $N$ free, we can
either fit low $z$ or high $z$ regions of the disk well, but not both
regions simultaneously; and (3) the two disk model is superior at
fitting the vertical profile at all latitudes.  We conclude that our
galaxies are best modeled by the superposition of two distinct
components with unique scale heights.  

While we already suspected the galaxies were composed of multiple
stellar components based on the observed vertical color gradients and
the analogous structures present in the Milky Way, we have now shown
that this conclusion can be derived from $R$-band images alone.  This
analysis does not preclude the existence of additional components
beyond the two disks considered here, however, although our data do
not obviously require them.  Our fits also do not demand that the two
components trace kinematically and chemically distinct stellar
populations that are directly analogous to the thick and thin disks of
the Milky Way.  On the other hand, when combined with the color
gradients observed in Paper II, the data are fairly suggestive of the
presence of two genuinely distinct components.  We will revisit this
issue further in \S\ref{sub_heights}.

Ideally, we could use a statistical goodness-of-fit test to show that
a second disk component is required when modeling edge-on galaxies. To
establish the need for a thick disk in UGC 7321, \citet{Matthews00}
used an $F$-test defined by
\begin{equation}
F=\frac{[\chi^2(1)-\chi^2(2)]/(p-k)}{\chi^2(2)/(n-p)}
\end{equation}
where $\chi^2(1)$ characterizes the single disk model with $k$ free
parameters while $\chi^2(2)$ characterizes the more complex model with
$p$ free parameters and $n$ total data points.  Comparing our two disk
models to single disk models with fixed $N$, the $F$-test favors
two disks at the 95\% level or higher confidence for 32 of the 34
galaxies, confirming that the two disk model is a better fit than a
single disk, as seen in Figure~\ref{two_components}.  Even if the index
$N$ is allowed to vary, the two disk model is still favored in 31 of
the 34 galaxies.  Although the $F$-test works well for
\citet{Matthews00} when fitting one-dimensional profiles, there are
several caveats we must note for our sample.  First, our models do not
necessarily minimize the formal $\chi^2$ value because of our inverse
weighting system.  Second, the $F$-test relies on the $\chi^2$
formalism and thus assumes all errors are random and Gaussian.  As we
noted in \S\ref{sec_fitting}, our residuals are definitely
non-Gaussian, and therefore the results of any $F$-test should be
viewed as suggestive but far from conclusive.

\subsection{Why Not a Halo?}\label{nohalo}

In addition to modeling the galaxies as a superposition of thick and
thin disks, we investigated models composed of a single disk and a
``stellar halo'' component as advocated by \citet{Zibetti04}.  For our
halo model, we used a generalized Hubble density distribution
\citep{Wu02} with the luminosity density
\begin{equation}
\mathcal{L}_{\mathrm{halo}}(r,z)=\frac{\mathcal{L}_{0
\mathrm{,halo}}}{\{1+[r^2+(z/q)^2]/r_0^2\}^{\gamma/2} }
\end{equation}
Viewed edge-on, this density distribution projects to the surface
brightness profile
\begin{equation}\label{halo_eq}
\Sigma_{\mathrm{halo}}(R,z)=\mathcal{L}_{0\mathrm{,halo}}
\sqrt{\pi}\frac{\Gamma[(\gamma-1)/2]}{\Gamma(\gamma/2)}\times
r_0^\gamma[r_0^2+R^2+(z/q)^2]^{(1-\gamma)/2}
\end{equation}
where $\Gamma$ is the standard gamma function

We find our data strongly prefers a second disk component to a halo.
Over half of our halo fits converged on very flattened halos ($q \le
0.45$), essentially reproducing the properties of a thick disk,
although one with a radial gradient in scale height.  In addition,
40\% of the fits converged to halo luminosities that are less than $1
\%$ of the disk luminosity, implying that the fitting procedure cannot
actually use the new component to improve upon the single disk fits.
When unconstrained, the halo exponential parameter $\gamma$ ran away
to very large or small values (producing a uniform background or a
compact point source), again implying that a power-law halo is not the
appropriate model for the light distribution at high z.

Using the $F$-test defined in \S\ref{sec_need2} to compare the two
disk fits to 9 free parameters with the disk plus halo fits with 10
free parameters, we found only 11 of the galaxies were better fit with
a halo than the second disk.  Even when $\chi^2$ suggested that the
halo model was a better fit, the flattening parameter converged to
extreme values (less than 0.4, or greater than 1) in 7 of the 11
cases, thus flattening the halo into a more disk-like structure.  In
those cases, the preference of a halo component may indicate the
presence of a radial gradient in disk scale height.  In the few cases
when a preferred halo fit remained roughly circular, it was because
the halo collapsed to fit a small central bulge or star-forming
region.  Because these regions are bright, they can greatly affect the
formal value of $\chi^2$ and the $F$-test will prefer the halo model
despite no real improvement in fitting the flux at large scale
heights.

The poor results of our attempted halo fitting do not explicitly rule
out the presence of a halo at lower surface brightnesses than we can
detect in our images.  Indeed, a stellar halo like the MW's would only
start to dominate the thick disk component at $\mu_R \sim 27.5$
mag/$\square$\arcsec ~\citep{Morrison97} around $z \sim 10z_0$ (our
fits extend to only $z\sim3-4z_0$).  In M31 the stellar halo
population dominates at a projected radius of $\sim30$ kpc and a
surface brightness level of $\mu_{V}\sim31$ mag/$\square$\arcsec
~\citep{Guh05}, and thus comprises $<5$\% of the total stellar
luminosity.  As before, if a comparable halo component was present in
our sample it would be much too faint to be detected in our data.

\citet{Zibetti04} fit a composite galaxy created by stacking over 1000
edge-on spirals from the Sloan Digital Sky Survey.  Using equation
\ref{halo_eq}, they found a slightly flattened halo with $q=0.50$ in
$g$, 0.60 in $r$ and $i$, and 0.70 in $z$.  There are several reasons
we believe the excess light we detect at high latitudes is not
equivalent to the halo component discussed by \citet{Zibetti04}: (1)
Our $R$-band is close to $r$ and $i$, yet when we try to fit a halo
component, our values of $q$ are much lower with a median value of
0.4; (2) The \citet{Zibetti04} halos only begin to dominate the
surface brightness at very large heights ($z=16 z_0$.)  beyond our
1-$\sigma$ cropping limit; (3) \citet{Zibetti04} find that their
stellar halo becomes prominent at a surface brightness of
$\mu_r\sim27$ mag/$\square$\arcsec, fainter than what we can detect in
our individual images.

%################################
%F-TEST Code in Final_plots/why22.pro

\subsection{Dust effects}\label{dust_effects}

\begin{figure}
%\plotone{plots/tf.eps}
\plotone{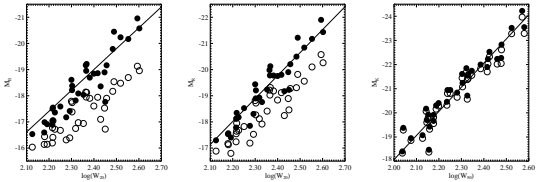}
\caption{ Tully Fisher relation derived from the single disk fits.
Open circles show points uncorrected for internal extinction while
solid circles have been corrected for internal extinction using the
method of \citet{Tully98}.  Solid lines on the left and middle panel
show TF relations from the Hubble Key Project \citep{Sakai00}.  The
solid line in the right panel shows the $K^\prime$ TF-relation of
\citet{Verheijen01}.  The luminosities for the two disk fits show a
comparable offset.  \label{tf}}
\end{figure}

We have previously found that galaxies in our sample with rotational
velocities greater than 120 km s$^{-1}$ host concentrated dust lanes
\citep{Dalcanton04}.  We therefore need to consider the effect that
dust extinction will have on our fitted parameters.  To quantify the
amount of extinction in our edge-on sample, we compare the total
luminosities of our best fit models to the Hubble Key Project
Tully-Fisher relation \citep{Sakai00} in Figure~\ref{tf}.  We find
that all of our galaxies, even those without recognizable dust lanes,
lie significantly below the TF-relation for face-on spirals.  If we
apply the extinction correction of \citet{Tully98}, however, our extinction
corrected total luminosities move nicely onto the face-on TF relation.

Their offset from the Tully-Fisher relation implies that our models do
not capture all the stellar flux from our galaxies.  There are several
possible ways dust could influence our fitted parameters to yield
lower than expected total luminosities: (1) the peak surface
brightnesses could be too low; (2) the scale lengths could be too
short; (3) the scale heights could be too small; (4) the vertical
profile could appear less peaked than it truly is (e.g., a sech$^2$
instead of an exponential); or (5) some combination of the above.

We can say with some certainty that the scale lengths do not appear to
be shortened by dust extinction.  If anything, Figure~\ref{single1}
shows that our scale lengths are larger than those measured in face-on
systems.  In a similar fashion, it is unlikely that our scale heights
are shortened greatly due to dust, as their bias is likely to have the
same sign as the scale lengths.  Moreover, our weighting scheme
de-emphasizes the midplane region, and our scale height fit is
therefore dominated by flux coming from high galactic latitudes.
Therefore, any dust distribution which is concentrated along the
midplane, or uniformly distributed though the galaxy, should have
little to no impact on our fitted value for the scale height.  Only a
truly pathological dust distribution, such as one having large amounts
of dust at high $z$ compared to the midplane, would skew our scale
height parameter to lower values.

Having eliminated biases in the scale height and scale length, we find
that dust extinction is most likely affecting our choice of vertical
profile and/or the fitted value of the peak surface brightness.
Unfortunately, there is a degeneracy between these two parameters which
could only be broken if we knew the intrinsic dust distributions in our
galaxies.  If the dust affects only the midplane region, then the
error could be confined to just the vertical distribution, while a
more diffuse and vertically uniform dust distribution would lower the
central surface brightness but not affect the vertical profile.

We conclude that while our galaxies display clear signs of internal
extinction caused by dust, the lost flux will cause us to either pick
the wrong vertical profile (which is not of particular importance
since we are not concerned with the midplane behavior), and/or
underestimate the overall flux normalization as parameterized by the
edge-on central surface brightness.  However, since the empirical
extinction correction of \citet{Tully98} does an excellent job moving
our galaxies onto the TF relation (despite the correction originally
not being intended for use on galaxies with extreme inclination
angles), we chose to apply this correction to the luminosity of our
thin disk component.  We do not assume any correction for the extended
thick component, since a much smaller fraction of its projected area
would be obscured by dust.

\subsection{Results of Thick + Thin Disk Fits}

We now discuss the results of fitting two disk components.  We fit a
total of six 2-disk models, each with different combinations of
$N=1$ and $N=2$ vertical profiles for the thick and thin components.
We also considered models convolved with a $\sigma=1$\arcsec ~circular
Gaussian (to model seeing) and models where the midplane ($\pm
z_{0,single}$) is masked (to avoid dust lane contamination).  The
properties of the models are described in Table~4.  We
use the inverse-weighting scheme for all the fits, as we found that
one disk component always collapses to fit bright midplane structures
if more conventional weighting is adopted.  As discussed above, we fit
the two-disk models only to the $R$-band images.  Our $K_s$-band data
does not go deep enough to reliably detect the thick disk component,
and the $B$-band suffers from dust extinction, is biased towards young
stellar populations, and is a poor tracer of the faint red light
expected from an old thick disk.

The resulting parameters for the fits are listed in
Table~5.  For the central value of each parameter
we list the median value of convergent $N=2$ models.  The
uncertainties are the full range of values to which the different
models in Table~5 converged, as discussed in
\S\ref{sec_fitting}.  We also list the ratio of total luminosities for
the model thick and thin disks.  The range of luminosity ratios
include models with disks having $N=1$ or $N=2$.  The luminosity
ratios are calculated only for flux which falls inside the 1-$\sigma$
noise region (i.e. only the region that was included in the fit).  The
true luminosity ratios could therefore be different from our quoted
values if the disks extend far beyond our detection limits.  We have
measured the size of this correction by extrapolating the fits and
find it can change the luminosity ratios by only 10\% at most.  The
luminosity ratios in Table~5 do not include the
extinction corrections derived in \S\ref{dust_effects}.

%##################RATIO OF SCALE HEIGHTS################
 \subsubsection{Scale heights of the thick and thin disks} \label{sub_heights}

\begin{figure}
%\plotone{plots/2d_z0s.eps}
\plotone{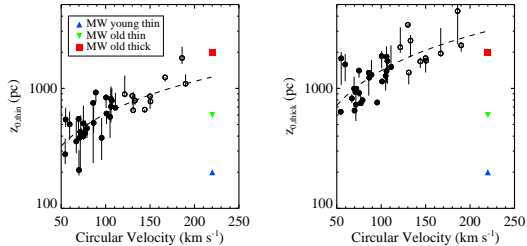}
\caption{Scale heights of thin and thick disks. Values for the Milky
Way from \citet{Larsen03} are plotted for comparison using $z_0=2h_z$
for an exponential vertical profile.  Open symbols are used for
galaxies with prominent dust lanes.  Dashed lines show power-law fits
to the data ($z_{0,\rm{thin}}=(610\mathrm{~pc}) (\frac{V_c}{100
\mathrm{~km~s}^{-1 }})^{0.90}$ and $z_{0,\rm{thick}}=(1400 \mathrm{~pc})(
\frac{V_c}{100 \mathrm{~km~s}^{-1}})^{1.0}$ ). In galaxies that have
strong dust lanes, the scale height of the thin disk is likely to be
biased towards larger values.
\label{2d_z0s} }
\end{figure}

The scale heights of our thick and thin disks are plotted in physical
units in Figure~\ref{2d_z0s}.  The scale heights of both the thin and
thick disks increase systematically with circular velocity.  Fitting
power laws to the relations, we find $z_{0,thin}=(610\mathrm{~pc})
(\frac{V_c}{100 \mathrm{~km~s}^{-1 }})^{0.90}$ and $z_{0,thick}=(1400
\mathrm{~pc})( \frac{V_c}{100 \mathrm{~km~s}^{-1}})^{1.0}$ with RMS
scatters of 30\% in both cases.  In general, the scale heights of the
two disks bracket the scale height derived for a single disk, as
expected.

For massive galaxies with large circular velocities ($V_c \gtrsim
170$ km/s), our derived value for the scale height of the thin disk is
2-3 times larger than the MW's thin disk.  However, these galaxies
have the most prominent dust lanes, which may substantially obscure
our view of the thin disk.  It is therefore likely that the scale
heights of the thin disk may be significantly overestimated in these
cases.  The plot of $z_{0,\rm{R}}/z_{0,\rm{K}}$ for the single disk
fits (Figure~\ref{single2}, lower right) is also consistent with this
interpretation.  Unfortunately, this limitation is unavoidable until
sufficiently deep $K$-band data is available.

\begin{figure}
%\plotone{plots/fig_zratio.eps}
\plotone{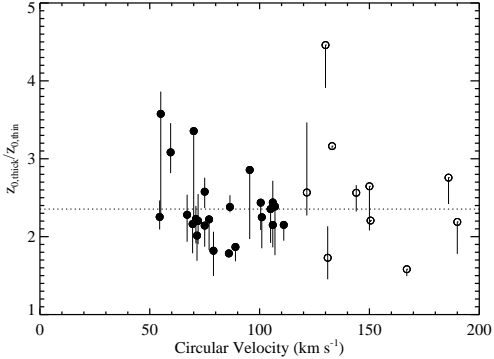}
\caption{Ratios of the scale heights for the thick and thin disks.
Error bars represent the full range of ratios to which different models
converged.  Galaxies with prominent dust lanes are plotted as open
circles.  The dotted line shows the median value of
$z_{0,thick}/z_{0,thin}=2.35$.\label{zrat}}
\end{figure}

\begin{figure}
%\plotone{plots/other_z.eps}
\plotone{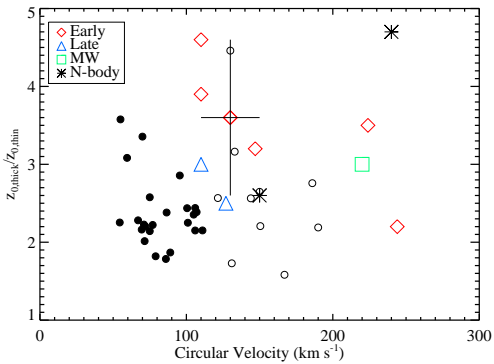}
\caption{Comparison of our scale height ratios to values drawn from
the literature. \label{other_z} We compare to early-type galaxies (Sb
and earlier; red diamonds), late type galaxies (Sc and Sd; blue
triangles), the MW (green square), and simulated galaxies.  The range
of values available for the \citet{Pohlen04} sample of early type
galaxies is plotted as a single point with error bars.}
\end{figure}

Figure~\ref{zrat} shows the ratio of the thick to the thin disk scale
height $z_{0,thick}/z_{0,thin}$.  We find a mean ratio of 2.5 with a
scatter of 30\%.  In Figure~\ref{other_z} we show our data along with
other thick and thin disk scale heights derived from the literature.
For the Milky Way, these scale heights are derived from star counts.
For the other literature values, the scale heights are derived either
from fitting double exponential profiles to 1-d cuts through the
galaxies or from 2-d fitting similar to the procedure used in this
paper.  We summarize these other results in Table~2.
Figure~\ref{other_z} indicates that our scale height ratios are
slightly lower than those measured in other systems
($z_{0,thick}/z_{0,thin}\sim3$), implying that our derived thick disks
may be $\sim25\%$ thinner and/or our thin disks are thicker than those
derived in other galaxies with other methods.  However, our median
$z_{0,thick}/z_{0,thin}$ is very similar to \citet{Neeser02}'s
measurement of the LSB galaxy ESO 342-017 , the most comparable galaxy
in the literature to galaxies in our sample.  These differences may
indicate that the thick disks of early type galaxies may be
proportionally thicker than those of late type galaxies.

\begin{deluxetable}{ l c c c c c c c l}
%\tabletypesize{\small \footnotesize \scriptsize}
\tabletypesize{\scriptsize}
%\rotate
\tablewidth{0pt}
%\tablenum{num}
%\tablecolumns{19}
%\tableheadfrac{num}
\tablecaption{Thick disks from the literature}
\tablehead{ 
\colhead{galaxy name} & \colhead{Type} & \colhead{Band} & \colhead{Fitted}  &\colhead{V$_{\mathrm{c}}$}(km s$^{-1}$)&
\colhead{$\frac{z_{0,thick}}{z_{0,thin}}$} & \colhead{$\frac{h_{R,thick}}{h_{R,thin}}$} 
 & \colhead{$\frac{\mathrm{L}_{thick}}{\mathrm{L}_{thin}}$}
&\colhead{Reference} } 
\startdata
34 galaxies & Sd  & $R$ &2-d& 55-190 &1.6-5.5 & 0.6-1.6 &  0.07-7 & This study \\
6 galaxies & Sd & star counts & 1-d& 67-131 & 1.7-2.7 & - & - &\citet{Seth05b} \\
ESO 342-017 & Scd & $R$ &1-d& 127 & 2.5 & 1 & 0.45 & \citet{Neeser02}\\
IC 5249 & Sd  & $R$ &1-d & 110 & 3 & 0.6 & - & \citet{Abe99} \\
MW & Sbc & star counts &2-d& 220 & 3 & 1.3 & $\sim 0.13$ & \citet{Larsen03} \\
NGC 6504 & Sb & $R$ & 1-d & 110\tablenotemark{1} & 3.9 & - & $\sim$0.4 & \citet{vanDokkum94} \\
NGC 891 & Sb & $R$ & 1-d &224 & 3.5 & - & 0.12 & \citet{Morrison97}\\
NGC 4565& Sb & 6660 \AA &1-d & 244& 2.2 & 1.4 & - & \citet{Wu02}\\
5 galaxies & S0 & $R$ and $V$ &2-d & $\sim130$\tablenotemark{2} & 2.6-5.3 & 1.7-1.9 & 0.33-1.0 & \citet{Pohlen04}\\
NGC4710	& S0 & $R$ &1-d &  147& 3.2	& -&- & \citet{deGrijs96} \\
NGC4762	& S0 & $R$ &1-d & 110& 4.6	& -& -& \citet{deGrijs96}\\
Simulation & - & - &-& 240 & 4.7 & - & 0.35 & \citet{Abadi203}\\
Simulation & - &- &-& 150 & 2.6 & 0.63 & 0.8 & \citet{Brook03}\\
\enddata
\tablenotetext{1}{Estimated from Tully-Fisher relation}
\tablenotetext{2}{Dynamical information only available for 2 of the 5 galaxies}
 \label{other_table}
\end{deluxetable}

We also do not {\emph{a priori}} know whether our thick and thin
components are strict analogs of any particular component in the disk
of the Milky Way, which is usually broken into at least three
components; (1) the ``young star-forming disk'' ($z_0\sim 200$ pc) which is
dominated by molecular clouds, dust, and massive OB stars; (2) the
``old thin disk'' ($z_0\sim 600$); and (3) the ``thick disk''
($z_0\sim 2$ kpc) \citep{Bahcall80, Reid93, Buser99, Larsen03, Ojha01,
Chen01}, which contains $\sim 15\%$ of the total disk light
\citep{Buser99, Chen01, Larsen03}.

One possibility is that our thin and thick disks might be analogous to
the MW's young star forming disk and old thin disk, respectively, with no
detectable analog of the MW thick disk.  However, we do not believe
the second component we have fit is an ``old-thin'' disk.  The
scale heights of our thin disks are larger than what has previously
been measured for thin star forming layers.  \citet{Matthews00} find
UGC 7321 has a ``young disk'' with $z_0 \approx 185$ pc.  Similarly,
IC 2531 has a young disk with $z_0\sim134$ pc \citep{Wainscoat89} and
the MW's young disk has $z_0\sim200$ pc \citep{Bahcall80,Reid93};
(using the conversion that at large scale heights $z_0\approx 2h_z$).
All known young star-forming disks thus have $z_0\sim200$ pc.  The only
galaxies in our sample that have thin disk scale heights approaching
values this small are far less massive than any of the galaxies in the
previous studies.

\begin{figure}
%\plotone{plots/2d_z0s_anil.eps}
\plotone{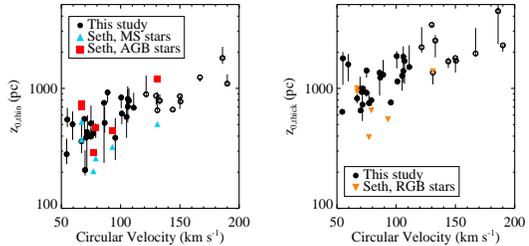}
\caption{Comparison of our results with the scale heights of different
stellar populations measured from resolved stars in 6 nearby galaxies
\citep{Seth05b}.  The component we have identified as the thin disk
appears to be intermediate between the scale height of young
Main Sequence stars and medium-age Asymptotic Giant Branch stars while
our thick disk component is similar to the old Red Giant Branch
populations.
\label{Anil_comp}}
\end{figure}

Having ruled out an old thin disk, we now consider the possibility
that our second thicker disk component is analogous to the MW's thick
disk.  We find strong support for this possibility from studies of
resolved stellar populations in similar systems
\citep[e.g.,][]{Seth05, Mould05, Tikhonov05}.  In particular, a recent
analysis of resolved stellar populations in edge-on galaxies by
\citet{Seth05b} separates stars into young Main Sequence (MS), older
Asymptotic Giant Branch (AGB), and still older Red Giant Branch (RGB)
stars.  \citet{Seth05b} find that the younger stellar populations have
systematically smaller scale heights than the ancient RGB population.
In Figure~\ref{Anil_comp}, we compare our thin and thick disk scale
heights with the MS, AGB, and RGB scale heights of \citet{Seth05b}.
We find that our thin disk components have scale heights very similar
to the young and intermediate age stellar populations of
\citet{Seth05b}, while our thick disk components have scale heights
similar to, or perhaps slightly larger than, the old RGB
populations. Figure~\ref{Anil_comp} supports that what we have labeled
the thin disk hosts a young and intermediate age stellar population
akin to the thin disk of the Milky Way while what we have labeled as
the thick disk traces a different older and redder population, not an
extension of the thin disk.  When coupled with our observation of
strong vertical color gradients (Paper II), and kinematic differences
above and at the midplane, we believe there is compelling evidence
that the second disk component required by our surface photometry does
represent a truly distinct stellar population.

%%##################RATIO OF SCALE LENGHTS################
\subsubsection{Ratio of scale lengths}\label{scale_length_sec}
\begin{figure}
%\plotone{plots/2d_hrs.eps}
\plotone{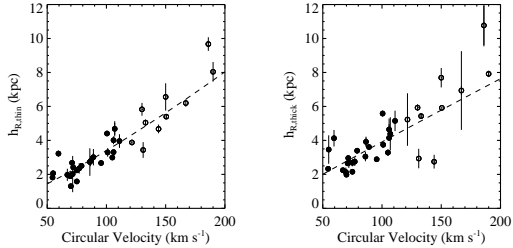}
\caption{Scale lengths of thick and thin disks from the 2-disk
fits. The dashed lines show power law fits.  Open symbols are used for
galaxies with prominent dust lanes.  Dashed lines show power-law fits
of $h_{R,\rm{thin}}=3.40(V_c/100\mathrm{~km~s}^{-1})^{1.2}$ kpc and
$h_{R,\mathrm{thick}}=3.9(V_c/100\mathrm{~km~s}^{-1})^{1.0}$ kpc.
\label{2d_hrs} }
\end{figure}

Physical values of the thick and thin disk scale lengths are plotted
in Figure~\ref{2d_hrs}.  We see a systematic increase in the radial
scale lengths of both disk components with galaxy mass.  The data are
well fit by $h_{R,thin}= (3.4 \mathrm{~kpc})
(\frac{V_c}{100\mathrm{~km~s}^{-1}})^{1.2}$ and $h_{R,thick}= (3.9
\mathrm{~kpc}) (\frac{V_c}{100\mathrm{~km~s}^{-1}})^{1.0}$ with RMS
scatters of 22\% and 29\% respectively.

\begin{figure}
%code to make in /net/grads-1/yoachim/condor/analy/range_fits.pro
%\plotone{plots/fig_rratio.eps}
\plotone{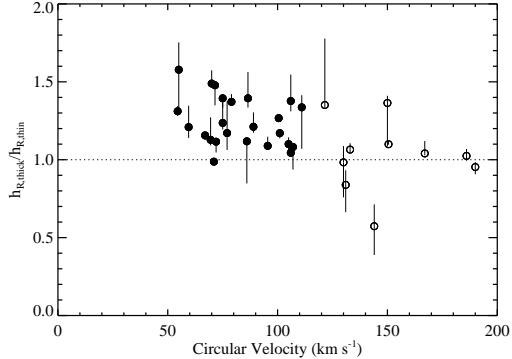}
\caption{Ratios of the scale lengths for the thick and thin disks.
The horizontal line indicates where the thin and thick disk components
have equal scale lengths.  Error bars represent the full range of
ratios to which different models converged, and are indicators of our
systematic errors.  Open symbols are used for galaxies with prominent
dust lanes.\label{rrat}}
\end{figure}

\begin{figure}
%\plotone{plots/other_hr.eps}
\plotone{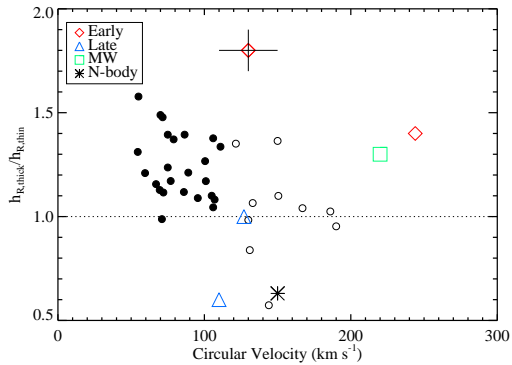}
\caption{Comparison of our scale height ratios to values drawn from
the literature.  Comparison points are the same as Figure
\ref{other_z}. \label{other_hr}}
\end{figure}

In Figure~\ref{rrat} we plot the ratio of the thick to thin disk scale
lengths.  We find that the thick disks have systematically larger
scale lengths for all but 5 galaxies.  Thick disks with long scale
lengths are in excellent agreement with previous thick disk
measurements, as shown in Figure~\ref{other_hr} where we include data
from the literature (Table~2).  In all but one
measurement of physical (i.e., non-simulated) thick disk scale
lengths, the thick disk is found to be slightly longer than the thin
disk.

We were initially concerned that this result could be a systematic
result of our weighting and masking scheme.  For example, if our
galactic disks truncate at large radii as found in other edge-on
systems \citep{Kregel02,Kregel04}, then our model fits would converge
to have the fainter thick disk dominate at large $R$.  However, we
included a model (Table~4) where the midplane is masked
(which also effectively removes regions of the galaxy where disk
truncation would be detectable) and still found that the thick disks
have longer scale lengths.

We note that there are some limitations in interpreting our scale
lengths, particularly for the thin disk.  First, the derived radial
scale lengths do not necessarily reflect the stellar radial scale
length.  The thin disk in particular shows a strong radial color
gradient, implying a mass-to-light ratio that decreases with
increasing radius (see color maps in Paper II).  This trend suggests
that the radial scale length of the stellar mass should be even
smaller for the thin disk, further increasing the ratio
$h_{R,thick}/h_{R,thin}$.  We may also have overestimated the scale
length of the thin disk if it is affected by dust in a manner similar
to the what is observed in our single disk fits (Figure~\ref{single1}
and \S\ref{dust_effects}).  Both of these effects suggest that
$h_{R,thick}/h_{R,thin}$ may be even larger than indicated by
Figure~\ref{rrat}.  On the other hand, HI is typically more extended
than the optical disk \citep{Swaters02I,Begum05}, such that the radial
scale length of the baryons in the thin disk may be longer than
indicated by $h_{R,\rm{thin}}$.

\subsubsection{Axial ratios of the thick and thin disks}

\begin{figure}
%\plotone{plots/single6.eps}
\plotone{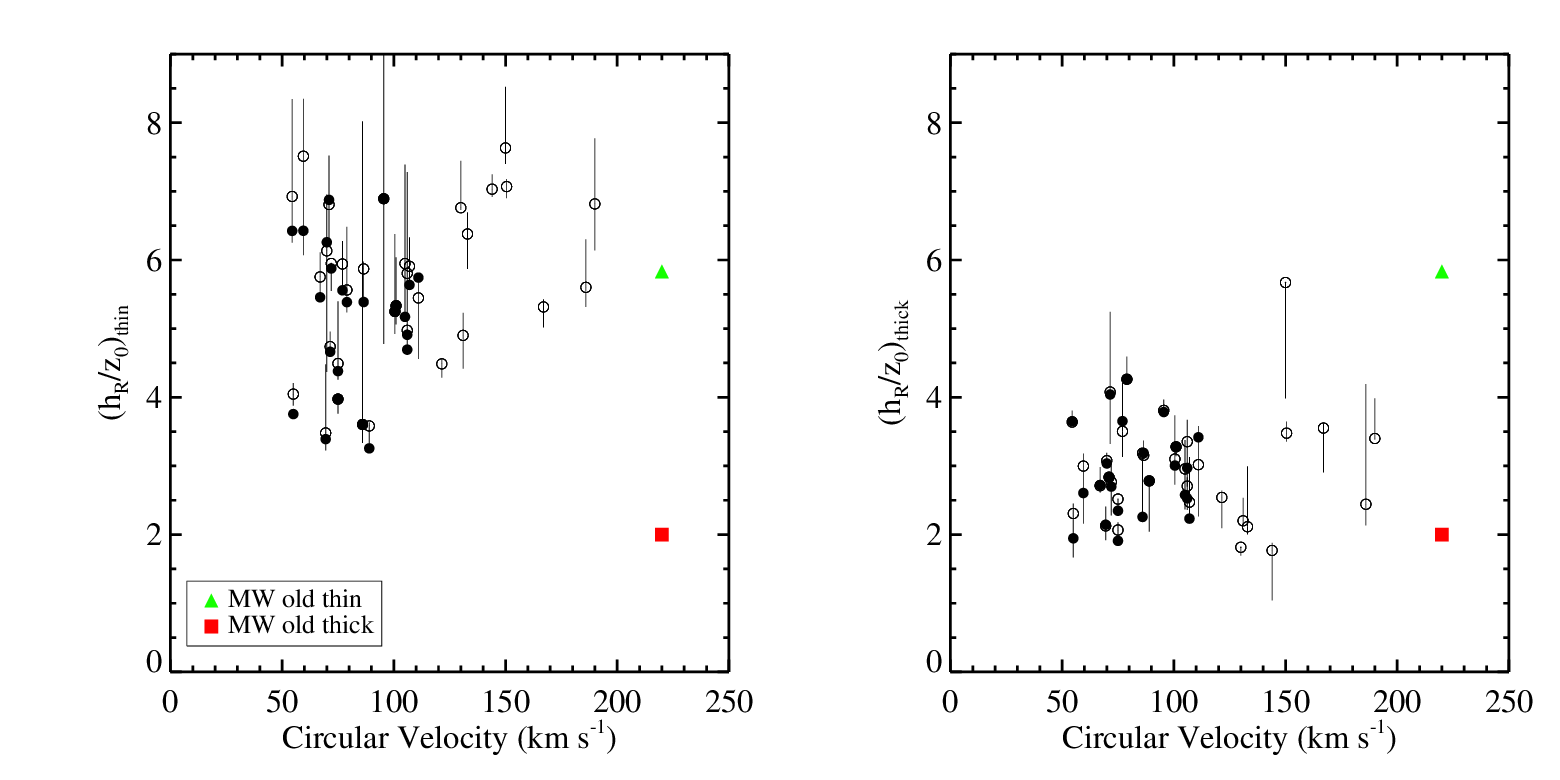}
\caption{ Axial ratio $(h_R/z_0)$ vs circular velocity for the thin
(left) and thick (right) disks.  Galaxies with prominent dust lanes
are plotted as open circles.  The axial ratios of Milky Way disk
components are plotted for comparison (Table~2).  The
axial ratios of our thin and thick disks agree well with the
comparable components for the MW.  \label{single6} }
\end{figure}

The axial ratios $(h_R/z_0)$ for our thick and thin disks are plotted
in Figure~\ref{single6} along with values for the MW thick and thin
components for comparison.  We find our thick disks have a mean
$h_R/z_0=3.4$ with an RMS scatter of 1.7 while the thin disk has a
mean value of $h_R/z_0=$4.7 and RMS scatter of 1.8.  Our thin disks
therefore tend to be comparable to the MW thin disk, but are slightly
rounder at low masses, in agreement with other studies (see
\S\ref{sec_sd}).  The axial ratios of the thick disks show a large
spread in axial ratios, and are in general comparable to, or slightly
thicker than the MW thick disk.  However, the radial scale length of
the MW is not well constrained since it is determined primarily from
star counts near the solar circle.  We also note that the thick disk
component is drastically rounder than the MW's old thin component,
further ruling out the old thin disk as an explanation for our second
disk component.

%######################Central Surface Brightnesses############
\subsubsection{Peak surface brightnesses}\label{csbs}

\begin{figure}
%\plotone{plots/2d_mus.eps}
\plotone{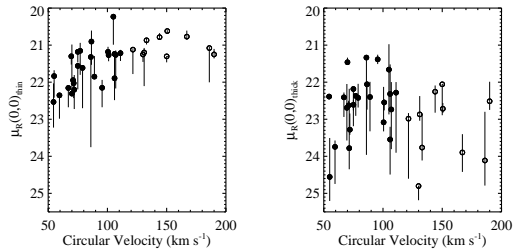}
\caption{Edge-on peak surface brightnesses for the 2-disk fits in
the $R$-band.  Galaxies with prominent dust lanes are plotted as open
circles.  \label{2dcsb} }
\end{figure}

The edge-on peak surface brightnesses for our two disk
components are plotted in Figure~\ref{2dcsb}.  There is a trend for
more massive thin disks to have brighter peaks, similar to the trend
seen in the single disk fits (Figure~\ref{single4}).  The thick disk
components show a large amount of scatter in their peak values. 

Performing a naive transformation to convert to the face-on
orientation, the central surface brightness becomes $\mu_0=
\mu(0,0)-2.5 \mathrm{log} (z_0/h_R)$.  We find that the average
central surface brightness of the thick disk is 0.6 \surfb fainter
than the thin disk, implying only $\sim35\%$ of the stellar flux in
the $R$-band would come from the thick disk if the galaxies were viewed
face-on.  In the more massive galaxies, the face-on central surface
brightness of the thick disk can be up to 2 \surfb ~fainter than the
thin disk.  These values do not include corrections made for
extinction. Presumably, the thin disk would suffer less extinction
when viewed face on, and would further dominate the observed stellar
flux.  It is therefore not surprising that the thick disk is largely
undetected in face-on galaxies.

The total integrated colors of the galaxies will be biased towards the
thin disk population as well.  After making the extinction corrections
in \S\ref{dust_effects}, we find the total integrated colors of our
low mass-galaxies ($V_c \lesssim 100 \kms$) are in the range $0.5 <
B-R < 1$, much bluer than the thick disk (see Figure~\ref{colors}
below).  Using the \citet{Bruzual03} stellar synthesis code, these
colors correspond to a stellar population burst with and age of
$\sim1$ Gyr, or a galaxy with a uniform star formation history.  Thus,
in spite of the substantial thick disk population, the mean colors of
the galaxy reflect only the youngest disk population.

\subsubsection{Ratio of luminosities}\label{lumrat_sec}
\begin{figure}
%\plotone{plots/lrat_excor.eps}
\plotone{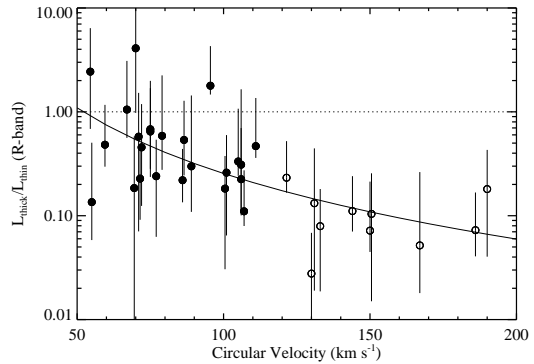}
\caption{Ratio of the total $R$-band luminosity of the thick disk
compared to the thin disk for the sample galaxies.  The thin disk
luminosities have been corrected for internal extinction.  The dotted
line indicates where the thick and thin disks contribute equally to
the total luminosity.  Error bars show the full range of values from
different models, and are indicative of our systematic
errors. Galaxies with prominent dust lanes are plotted as open
circles.  The solid line is a fitted power-law of the form
$L_{\rm{thick}}/L_{\rm{thin}}=
0.25(V_c/100\mathrm{~km~s}^{-1})^{-2.1}$.  \label{ext_cor} }
\end{figure}

We now compare the total luminosities of the thick and thin disks
(Figure~\ref{ext_cor}).  In our raw fits, the luminosity of the thin
disk is almost certainly underestimated due to the effects of dust, as
shown in \S\ref{dust_effects}.  To correct for dust, we assume that
all flux lost from extinction (Figure~\ref{tf}) should be assigned
to the thin disk.  This correction will give us the most conservative
estimate for the contribution of the thick disk to the total stellar
luminosity.  Figure~\ref{ext_cor} shows a strong trend with mass
(Spearman $\rho=-0.70$, 4.0$\sigma$).  Thick disks of high mass
galaxies ($V_c > 120$ km s$^{-1}$) contribute $\sim10$\% of the total
luminosity of the galaxy, while in lower mass systems the thick disk
contributes up to 40\% of the total luminosity.  This trend can be well
represented by the relation $L_{\mathrm{thick}}/
L_{\mathrm{thin}}=0.25 (V_c/100\mathrm{~km~s}^{-1})^ {-2.1}$, shown as
a solid line in Figure~\ref{ext_cor}.

\begin{figure}
%\plotone{plots/other_l.eps}
\plotone{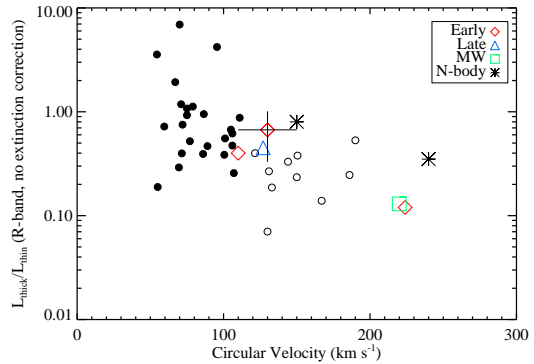}
\caption{Comparison of our luminosity ratios to others in the
literature. Comparison points are the same as Figure~\ref{other_z}.  Unlike
Figure~\ref{ext_cor}, we have made no correction of internal
extinction to allow for easier comparison with previous studies.
\label{other_l}}
\end{figure}

We compare our measurements to previous thick disk measurements in
Figure~\ref{other_l}.  Unfortunately, there are few measurements of
total disk luminosities in the literature.  When possible, we have
taken other authors' disk parameters and calculated the resulting total
luminosities (see Table~2).  For the Milky Way, the
local stellar density of thick disk stars has consistently been
measured between 4 and 10\% of the local thin disk density
\citep[e.g., ][]{Buser99, Chen01}, which corresponds to a total
luminosity ratio of $\sim$13\% for reasonable estimates of scale
heights, lengths, and mass-to-light ratios for the two disks.  Because
the values of $L_{\mathrm{thick}}/L_{\mathrm{thin}}$ from the
literature do not include internal extinction corrections, we compare
them to our uncorrected luminosity ratios.

%%%%%XXX--too many floats running around
%\clearpage

Figure~\ref{other_l} shows that our luminosities compare well with
other thick-thin disk systems in the literature.  The higher mass
galaxies in our sample ($V_{c}\sim 140-200$ km s$^{-1}$) tend to be
thin disk dominated with $L_{\mathrm{thick}}/L_{\mathrm{thin}} \sim
0.1-0.2$ (corrected for extinction), like the Milky Way and NGC 891.
Intermediate mass galaxies ($70 < V_{c} < 100 $ km s$^{-1}$) have
slightly more luminous disks, similar to measurements of ESO 342-017
\citep{Neeser02} and S0's \citep{Pohlen04}.  Unfortunately, we cannot
find any comparable measurements of thick disks in the low mass
systems ($50 < V_{c} < 70 $ km s$^{-1}$) that are thick disk dominated
in our sample.

We believe our measurement of the total luminosities are more robust
than the measures of the peak surface brightness .  Central surface
brightnesses depend strongly on the vertical profile and can vary
greatly from author to author.  On the other hand, our fits of the
total luminosity are good matches the the data
($|$m$_{\mathrm{model}}$ - m$_{\mathrm{observed}}|\sim0.2$ mags), and
fall on the Tully-Fisher relation (Figure~\ref{tf}).

\subsubsection{Mass ratios}\label{massrat_sec}

Figure~\ref{ext_cor} indicates that thick disk stars provide a
significant fraction of a galaxy's total luminosity.  However, as seen
in Paper II, the thick disk tends to have a redder color than the thin
disk, especially in low mass galaxies, and thus will have larger stellar 
mass-to-light ratios than the thin disk.  Therefore, the thick disk
may well dominate the stellar mass in many of our galaxies.  We
estimated the stellar disks' masses using the luminosities of the two
disk components, along with color information from our single disk
fits.  Specifically we used the spectrophotometric galaxy evolution
analysis of \citet{Bell01} to convert our $B-R$ color maps into
stellar mass-to-light ratios for each disk, and then convert our
luminosity ratios into mass ratios for the thick and thin components.

The initial analysis of vertical color gradients in our sample (Paper
II) suggested that the colors of thin disks vary systematically with
galaxy mass, but that the colors of thick disks are fairly uniform.
We therefore assumed that the thick disks have uniform colors and
mass-to-light ratios in each galaxy.  To convert disk colors to
masses, we first analyzed our $R$-band 2-disk fits to find regions
where the thick disk contributes more than 75\% of the total flux
inside the 1-$\sigma$ noise contour.  Out of 34 galaxies, 27 have a
clearly thick disk dominated region.  We created a model $B-R$ color
map of each galaxy using our 2-disk $R$-band model and single-disk
$B$-band fits.  Using the model images allowed us to avoid dust lanes,
HII regions, and foreground objects that would complicate an analysis
on the real images.  We then used this color map to find the average
$B-R$ value in the thick disk dominated regions and took that as the
approximate color for all the thick disk stars.  We assumed the thick
disk has a constant color, thereby guaranteeing that its structural
parameters will be the same in both the $B$ and $R$ bands.  With the
$B-R$ thick disk color from the model color map and the thick disk
structural parameters from the 2-disk fit, we then made a model $B-R$
color map for the thin disk by subtracting off the thick disk
component from both the $B$-band and $R$-band models.

We applied internal extinction and reddening corrections to our models
using the results of \S\ref{dust_effects} and assumed that dust had a
uniform effect on the thin disk colors but a negligible effect
on the thick disk.  Using this approximation, we found $E(B-R)\sim0.1$
for low mass galaxies and $\sim0.4$ for higher mass galaxies.
Although it is only a rough approximation, our reddening correction is
in good agreement with the radiative transfer model of
\citet{Matthews01} who find that most disk light in their modeled
edge-on galaxies suffer reddening of order $E(B-R) \sim 0.1$, and that
the reddening saturates at $E(B-R)=0.31$.

\begin{figure}
%\plotone{plots/disk_colors.eps}
\plotone{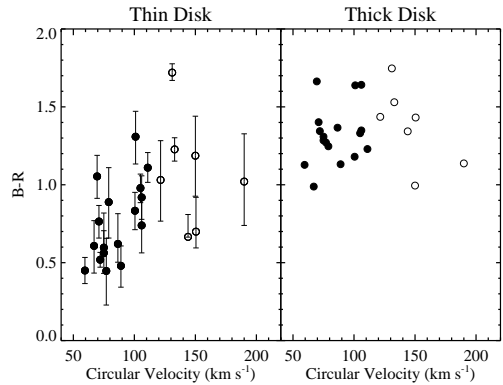}
\caption{Extracted colors for the thick and thin disks. The thin disk
colors shows the full range of $B-R$ values for the midplane between
$h_R< R < 3 h_R $.  The thin disk has been corrected for internal
extinction, but we assume no correction for the thick disk.  Open
circles show galaxies with dust lanes.  \label{colors}}
\end{figure}

\begin{figure}
%\plotone{plots/col_comp.eps}
\plotone{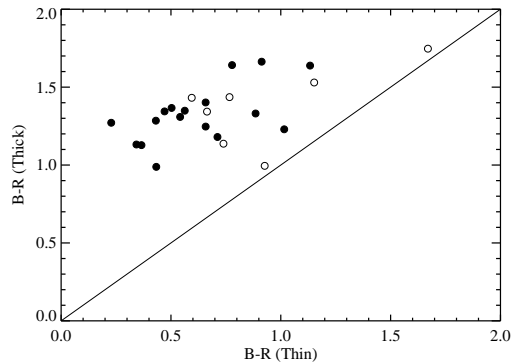}
\caption{Comparison of thin and thick disk colors.  The diagonal line
indicates where the two components have equal color.  Thin disk colors
have been corrected for internal extinction.  Open circles show
galaxies with dust lanes.  \label{col_comp}}
\end{figure}

The resulting colors for thick and thin disks are plotted in
Figure~\ref{colors}.  The thick disks tend to be red with $1.0
\lesssim B-R \lesssim 1.7$, while the thin disks are blue in low mass
galaxies and become nearly as red as the thick disks in the higher
mass galaxies.  This trend is also seen in Figure~\ref{col_comp},
where we directly compare the colors of each component.

Using the colors shown in Figure~\ref{colors}, and the color dependent
stellar mass-to-light ratios from \citet{Bell01}, we converted the
thick and thin disk luminosities to stellar masses using $M=(M/L)_R
L_R$, where $L_R$ is the extinction corrected $R$-band luminosity from
\S\ref{lumrat_sec}.  For the thin disk, we calculated $(M/L)_R$ using
the \citet{Bell01} model which assumes a Salpeter IMF and metallicity
of $Z=0.02$, for the thick disk we use the same model with $Z=0.08$.
Overall, our results were insensitive to the stellar evolution and
metallicity differences covered in the \citet{Bell01}
models. 

\begin{figure}
%\plotone{plots/disk_masses_ex.eps}
\plotone{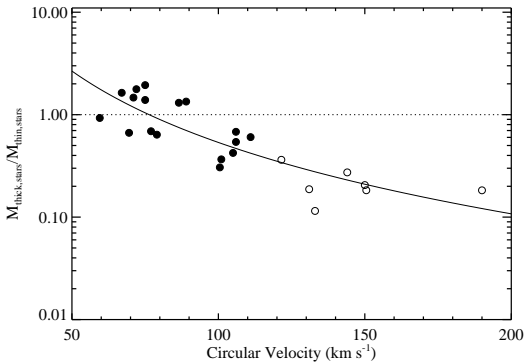}
\caption{Stellar mass ratios of the thick and thin disks.  The thin
disks luminosities and corresponding masses have been increased to
account for dust extinction. Galaxies with prominent dust lanes are
plotted as open circles.  The solid line is a power-law fit
$(M_{\mathrm{thick}}/M_{\mathrm{thin}})_{\mathrm{stars}}
=0.53(V_c/\mathrm{100~km~s}^{-1} )^{-2.3}$.
\label{massrat} }
\end{figure}

The resulting mass ratios of the thick and thin disks are plotted in
Figure~\ref{massrat}.  As expected, Figure~\ref{massrat} confirms the
features from our luminosity analysis.  First, there is a strong trend
for lower mass galaxies to have a larger fraction of their stellar
mass in a thick component.  The trend has a Spearman correlation of
$\rho=-0.86$ ($4.1\sigma$) and can be well fit with the relation
$M_{\mathrm{thick}}/M_{\mathrm{thin}} = 0.53(V_c/
\mathrm{100~km~s}^{-1} )^{-2.3}$.  Second, in low mass galaxies,
$\sim1/3$ to greater than 1/2 of the stellar mass is in the thick
disk.  Thus, the stellar mass of very low mass galaxies are dominated
by thick disk stars.

Part of the trend in Figure~\ref{massrat} may be due to low star
formation efficiency in lower mass disks.  These systems have high gas
mass fractions, and thus may not yet have built up a significant
stellar mass in the thin disk.  To investigate this possibility, we
calculated the baryonic mass fraction of the thick and thin disks,
assuming that all gas in the galaxies is associated with the thin disk
and that the thick disk is entirely stellar. We calculate the gas mass
as $M_{\mathrm{HI}}/M_\odot=236d^2\int S \mathrm{d}V$ where $d$ is the
distance to the source in Mpc, and $S$ is the flux density in mJy over
the profile width d$V$ in km s$^{-1}$ \citep{Zwaan97}.  To
account for He and metals, we make the standard correction
$M_{\rm{gas}}=1.4M\rm{_{HI}}$.  We do not include a correction for
molecular gas.  

\begin{figure}
%\plotone{plots/mass_rats_pgas.eps}
\plotone{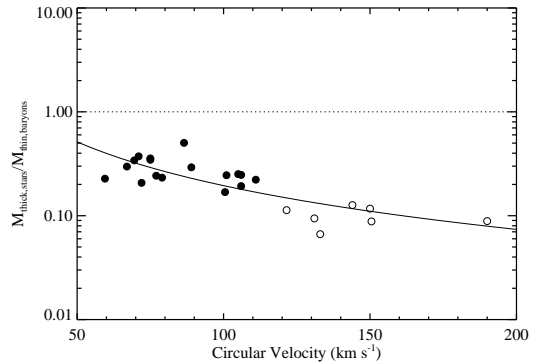}
\caption{Baryonic mass ratios of the thick and thin disks.  The thin
disks luminosities, and corresponding masses, have been increased to
account for extinction effects and include the estimated thin disk
mass stored in gas. Galaxies with prominent dust lanes are plotted as
open circles. The solid line is a power-law fit
$M_{\mathrm{thick,stars}}/M_{\mathrm{thin,baryons}}=0.19(V_c/\mathrm{100~km~
s}^{-1} )^{-1.4}$.\label{rat_w_gas}}
\end{figure}

\begin{figure}
%\plotone{plots/mass_fill.eps}
\plotone{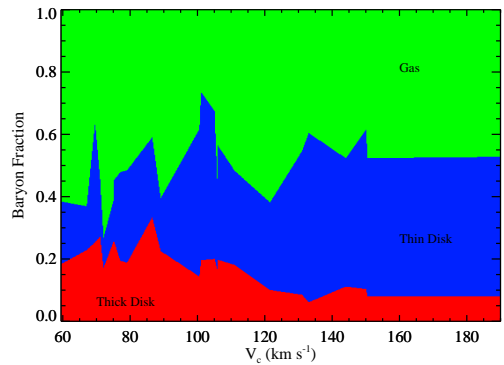}
\caption{Final baryon fractions of the thick and thin components as a
function of maximum velocity.  The thin disk has been corrected for
internal dust extinction. The baryonic mass fraction of the thick disk
clearly increases towards lower galaxy mass.  \label{mass_fill}}
\end{figure}

Figure~\ref{rat_w_gas} shows the resulting baryonic mass ratio of
thick and thin disks with the mass of HI gas added to the thin disk
component.  When the gas is included in the thin disk component, we
find that none of the galaxies remain thick disk dominated although
the baryon mass fraction in the thick disk does remain substantial for
low mass galaxies.  Eleven of our galaxies had no HI data and their
gas fraction was estimated by fitting a simple power law to the gas
fraction of our galaxies with HI measurements.  Figure~\ref{mass_fill}
shows the final calculated baryon fractions for all of the stellar and
gaseous components, and clearly indicates the increasing importance of
the thick disk in lower mass galaxies.

\section{The Formation of the Thick Disk}

Given evidence from the Milky Way and from nearby resolved galaxies,
we will assume in the following discussion that thick disks have a
formation mechanism distinct from that which forms the thin disk.  We
will also assume that the properties of ``thick'' and ``thin''
components from our 2-D fits will roughly approximate the properties
of the chemically and kinematically distinct thick and thin disk
analogs of the Milky Way.  As we have argued above in
\S\ref{sub_heights}, the structural properties of the fits are
consistent with those of the corresponding components of the Milky Way
and with the results of detailed stellar population studies in nearby
galaxies (e.g.\ Figure~\ref{Anil_comp}).  We therefore will simply
assume a perfect correspondence between our fits and distinct thick
and thin components for the rest of this discussion.  While this
assumption is not ideal, it is unavoidable, given that a full
kinematic and chemical analysis of the stellar components is
essentially impossible far outside the Local Group.

\subsection{The Merger Origin of Thick Disks} \label{mergersec}

As discussed above in \S\ref{intro}, there are three general classes
of thick disk formation scenarios -- one where the thick disks stars
form {\emph{in situ}}, one where the thick disk stars form in a thin
disk but are then impulsively ``heated'' to large scale heights, and
one where thick disk stars form first in other galaxies but are then
directly accreted.  In the last decade there has been a growing body
of evidence in favor of the latter two scenarios.  This evidence
includes the detection of strong kinematic differences between thick
and thin disks in other galaxies \citep{Yoachim05} and in the Milky
Way \citep{Gilmore02}, as well as evidence from chemical abundance
studies for extended star formation histories of Milky Way thick disk
stars \citep[e.g.,][]{Bensby05}.  These latter two scenarios are also
naturally accommodated in current theories of hierarchical structure
formation, where mergers and accretion are common.

Of the two merger-driven scenarios, we believe that the data favor an
accretion origin for thick disk stars.  The strongest evidence comes
from our previous measurements of thick and thin disk kinematics in
two late-type disks.  We found that thick disk stars are rotating with
only a small fraction of the rotational velocity of thin disk stars,
and are indeed counter-rotating in one of the two cases studied.  In
contrast to the observed behavior, simulations show that disks heated
by satellites would have nearly the same angular momentum as the
initial thin disk.  Therefore, barring the unlikely possibility that
the thin disk reforms from subsequent accretion of gas with angular
momentum opposite to the original disk, creating counter-rotating or
slowly-rotating thick disks via vertical heating alone is problematic.
Additional supporting evidence comes from recent HST studies of
resolved stars in nearby edge-on galaxies \citep{Seth05b,Mould05}.
While \citet{Seth05b} finds evidence for some {\emph{steady}} vertical
heating, the oldest population of RGB stars shows no evidence of the
vertical metallicity gradient expected if they were dominated by stars
that had participated in the steady heating.  Instead, the thick
population of RGB stars must have been established early, when the
merger rate was still extremely high.  At these early times, the
hypothetical picture of a well-defined stable thin disk impulsively
heated by a single merging event seems inappropriate.  In summary,
while the evidence does not yet conclusively rule out a single, thin
disk heating merger event as the origin of thick disk stars, we
consider it to be sufficiently unlikely that we will focus on
interpreting the results in this paper in terms of the accretion
scenario.

The idea that stars well above the Galaxy's midplane may have been
directly accreted from satellites was first discussed by
\citet{Statler88}\footnote{Statler's work refers to explaining the
kinematics of ``halo'' stars, but given the current understanding of
Galactic structure, the specific kinematic data he was trying to
explain were measurements of what we would now call thick disk
stars.}, and then codified as a distinct formation mechanism for the
thick disk shortly thereafter in the review article by
\citet{Gilmore89} and as ``Model 7'' in \citet{Majewski93}'s review.
Recent detailed studies of stellar structures in the Milky Way and M31
find evidence that stars are regularly accreted by massive galaxies.
For example, \citet{Martin04} find asymmetries in the distribution of
M-giant stars \citep[e.g., the Galactic 'Ring'][]{Newberg02,Yanny03}
that are well explained by a single dwarf galaxy accretion event.
\citet{Martin04} also note that their modeled accretion event results
in accreted stars having orbits similar to thick disk stars, and that
the thick disk may be continually growing through in-plane accretion
of dwarf galaxies.  M31 also appears to also be actively disrupting
dwarf galaxies with an observable stellar stream \citep{Ibata04} and a
large extended disk \citep{Ibata05}.

Direct accretion as the dominant origin of thick disk stars has
recently been revived by several numerical studies of disk galaxy
formation.  Analyzing an N-body simulation of a moderately low mass
spiral formed in a cosmological context, \citet{Abadi203} found a
well-populated thick disk, more than half of which was made up of
stars originally formed in accreted satellite galaxies.  Although by
no means definitive due to the simulated galaxy's unrealistically
large bulge, \citet{Abadi203}'s work lead to a revival of the notion
that thick disk stars may have formed outside their host galaxies.
Subsequent simulations of a collapsing sphere seeded with
perturbations by Brook et al. (2004, 2005) \nocite{Brook05} generated
thick disks associated with an early episode of chaotic merging.
Unlike \citet{Abadi203}, \citet{Brook04} argued that the thick disk
stars formed {\emph{in situ}} from large velocity dispersion gas
deposited by the satellites as they merged together to the final disk
structure.  Both simulations have significant limitations, making it
impossible to decide in favor of either scenario at this time, but
both stress the importance of merging and accretion in setting
properties of the thick disk.

In the context of hierarchical galaxy assembly, the above simulations
point to a straightforward picture of disk formation that necessarily
leads to the formation of thick disks.  At high redshift, galaxies
exist largely as a collection of sub-galactic fragments.  These
fragments consist of gravitationally bound dark matter ``mini-halos'',
many of which presumably host some amount of baryonic material.
Because these systems are high in the merging hierarchy, they would be
expected to be relatively dense, and thus some of the gas hosted by
these sub-units is likely to form stars.  Early on, the merging rate
will be very high, and as these fragments come together, their orbits
will tend to circularize, align, and decay due to dynamical friction,
as in \citet{Statler88} and early simulations by \citet{Quinn86} and
\citet{Walker96}.  The merged subunits will form a rotating flattened
structure provided that the net angular momenta of the satellites is
sufficiently high.  When the merging rate declines sharply ($z\sim 3$)
\citep{Zhao03}, the disk will be left in place as a long-lived
structure relatively unperturbed by significant accretion events.  Any
dense gas associated with the pre-galactic fragments must then either
form stars in a burst during the final merger of the fragments, as in
the \citet{Brook04} simulations, or cool into a thin disk which later
converts into stars.

Within this scenario, {\emph{any}} stars that formed in the sub-units
and were not tidally stripped at large radii must necessarily wind up
in a thickened disk structure, with a vertical velocity dispersion
equal to or greater than the velocity dispersion of the typical
pre-galactic fragment.  Because they are effectively collisionless,
the accreted stars cannot lose energy and cool into a thinner disk,
and must retain a large fraction of the initial velocity dispersion
and angular momentum of the satellite in which they formed.  With this
in mind, it seems impossible to imagine {\emph{not}} forming a thick
disk \citep[unless star formation was completely suppressed at early times,
for example by reionization, e.g., ][]{Bullock00, Gnedin00}. The only other
possible destination for the accreted stars would be the bulge or
stellar halo.  However, the sample considered here is essentially
bulgeless.  We also find no evidence for a luminous stellar halo down
to our limiting surface brightness.  Taking a conservative estimate
for the surface brightness of the brightest stellar halo that could be
present, but undetected in our data, we find that any stellar halo
must be less than 15\% of the luminosity of the thick disk.  This
estimate suggests that the majority of directly accreted stars settle
into the thick disk.

In addition to the theoretical arguments for forming thick disks via
direct accretion of stars, there is a slowly growing body of
observational evidence for this process seen {\emph{in situ}} at high
redshift.  First is the analysis of high redshift ``clump-cluster''
galaxies by \citet{Elmegreen05}.  Morphologically, these galaxies
appear to consist of many distinct, high surface brightness clumps
merging together.  \citet{Elmegreen05} argue persuasively that these
systems will wind up in a thickened disk with high velocity
dispersion, and are thus likely precursors to thick disks.  The colors
of the clumps suggest that they already contain some stars, and are
not pure gas systems.  \citet{Elmegreen05} also find field
counterparts of the clumps, suggesting that some of the stars may have
formed before being accreted into the galaxy.  The second piece of
evidence is the kinematic study of \citet{Erb04}, who find that lumpy
disk-like structures at $z\!\sim\!2$ show little net rotation.  If
their sample consists primarily of the edge-on counterparts of the
galaxies in the \citet{Elmegreen05} study, then the lack of strong
rotation would be consistent with what is expected for material that
forms a thick disk.  Although this study traces H$\alpha$ kinematics
only, and thus leaves the kinematic state of any associated stars
unconstrained, it would be peculiar if any stars associated with the
accreting gas did not show similarly perturbed kinematics.

\subsection{Constraints from the Structures of Thick and Thin Disks}

In the above accretion scenario, the properties of the thick disk are
fixed primarily by the stellar content and orbital properties of the
pre-galactic fragments which merge to form the final stable disk.  The
properties of the thin and thick disks are then set by the kinematics
and gas mass fractions of the pre-galactic fragments when they merge.

Within this scenario, we now discuss the implications of three
significant properties of thick disks uncovered by our data: first,
that thick disks are a ubiquitous and necessary component in modeling
late-type edge-on galaxies; second, that the stellar mass of the thick
disk is increasingly dominant in lower mass galaxies; and third, that
thick disks have systematically larger radial scale lengths than thin
disks.

\subsubsection{The Ubiquity of Thick Disks}

The 2-D fits of our sample confirm the initial suggestion of
\citet{Dalcanton02} that thick disks are a ubiquitous component of
disk galaxies.  Essentially all (32 of 34) of the galaxies which were
suitable for fitting were statistically significantly better fit by a
second disk component (e.g.\ Figure~\ref{two_components}).  This
result adds to the published detections of thick disks in 
earlier type systems (summarized in Table 2).  Thick disks are now
routinely discovered in every galaxy that has been searched for
them\footnote{The one exception is NGC 4244, which \citet{Fry99}
analyzed using fits to 1-D cuts of the vertical $R$-band light
distribution.  Based on the lack of a clear break in the vertical
surface brightness profile, \citet{Fry99} claimed there was no thick
disk in this galaxy.  However, subsequent analyses of the resolved
stellar population in NGC 4244 by \citet{Seth05b} and
\citet{Tikhonov05} revealed the presence of a clear extra-planar
population dominated by old red giant branch stars, whose global
distribution was characteristic of a thick disk.}.

The evidence therefore supports the idea that thick disks are a
generic property of all galaxies with disks, from S0's to Sm's, from
high masses ($V_c\!\sim\!250\kms$) to low ($V_c\!\sim\!50\kms$).
Thick disks must therefore be a natural by-product of disk galaxy
formation, independent of the formation of a bulge.  The ubiquity of
thick disks can be easily explained if most thick disk stars are
directly accreted from pre-galactic fragments.  As we argue above, if
{\emph{any}} star formation has taken place in the fragments, some
fraction of those stars must wind up in a thick disk.  The only way to
avoid depositing the stars in a thick disk would be if the fragments
were completely tidally disrupted at large distances from the central
galaxies.  However, at large distances the matter density should be
much lower than in the dense cores of the low mass galactic fragments,
making it unlikely that every merging satellite would experience
complete disruption.

The existence of widespread thick disks also suggests that there has
been ample star formation in the very low mass halos which merge
together to form larger galaxies.  Most sub-units must have
established stellar populations before merging.  If instead the
sub-units were entirely gaseous, disk galaxies would have only a thin
disk component.  Thus, there cannot have been total suppression of
star formation by reionization up until the epoch of thick disk
formation.  

Finally, the pervasiveness of thick disks also presents an additional
problem for merger heating scenarios.  It seems unlikely that every
galaxy in our sample would have had both a merger that created a thick
disk and accretion that reformed a thin disk.  If merger heating was the
primary driver of thick disk formation we would expect to find some
galaxies that were able to avoid a destructive merger, or that failed to
subsequently reform a thin disk.  Instead, all of our galaxies require
both thin and thick disk components.

\subsubsection{The Increasing Importance of Thick Disks in Lower Mass 
Galaxies}

In the merging picture we have adopted, sub-galactic fragments
contribute both stars and gas to the final galaxy.  The stars wind up
in the thick disk, and the gas settles into the thin disk, where it
gradually converts into stars.  From Figure~\ref{mass_fill} we see
that low mass disk galaxies have roughly 25\% of their baryonic mass
locked up into thick disk stars, while massive galaxies have only
10\%.  Figure~\ref{mass_fill} therefore implies a systematic variation
in the gas richness of sub-galactic fragments at the time disks
coalesce.  In massive late-type galaxies, 90\% of the baryonic mass
must have remained gaseous during disk assembly, while in low mass
galaxies only 75\% had not yet converted to stars.

Note that while we are stressing the accretion of stellar material to
form the thick disk, our results prove that the vast majority
(75-90\%) of baryonic accretion must have been gaseous.  If some
fraction of thick disk stars did form {\emph{in situ}} as suggested by
Brook, then the fraction of gaseous accretion must have been even
higher.

There are several ways that lower gas mass fractions in the precursors
of low mass galaxies may be achieved.  One possibility is that the
transformation of gas into stars proceeded further by the time the low
mass disk galaxies coalesced.  This more complete transformation in
low mass disks could be due either to a later epoch of assembly, or to
higher gas densities and thus higher star formation rates in the
precursor clumps.  However, in a closed box model, the resulting thick
disk stars would have higher metallicities.  In contrast, the
estimates of the metallicities of extra-planar, RGB stars in
\citet{Seth05b} suggest that the metallicities of the thick disk stars
are systematically lower in lower mass galaxies, compared to the Milky
Way.  We therefore rule out the possibility that star formation was
more ``complete'' in the precursors of lower mass galaxies.

Supernova feedback is an alternative pathway to the preponderance of
less gas rich sub-units in low mass galaxies.  Much of the disk material
was initially in several subunits that were necessarily of lower mass
than the final galaxy.  Thus, the merging fragments must have had
lower escape velocities, allowing supernova-driven winds to more
effectively drive gas and metals from the galaxy at this early stage.
The increased efficiency of SN winds in the sub-units would
simultaneously decrease the gas mass fractions and maintain low
metallicities in thick disk stars in low mass galaxies\footnote{Note,
however, that the overall gas mass fraction of low mass galaxies can
remain high to the present day.  The disks of lower mass galaxies have
systematically lower baryon surface densities \citep[e.g., ][]{Swaters02I,
Hunter04}, and thus are inefficient at converting gas into stars.
Their low star formation rate thus allows them to have higher gas mass
fractions today, even though they were comparatively gas poor at the
time their disks were assembled.}.

We can estimate the amount of gas loss needed to produce the observed
trends as follows.  First, we assume that the observed baryon fraction
in the stellar thin disk and gas component of massive galaxies
($\sim\!90$\%, Figure~\ref{mass_fill}) is indicative of the gas to
stellar mass fraction in subgalactic fragments that are too massive to
experience significant SN blowout.  We then assume that the precursors
of lower mass galaxies lose enough gas to bring their gas to stellar
mass fraction down to $\sim\!75$\% at the time of disk assembly.
These simple assumptions imply that the sub-units of low mass galaxies
must have lost 60\% of their initial baryonic mass.

Because we have ignored possible tidal stripping of stars during
galaxy assembly, the actual amount of gas lost from the precursors of
low mass disks may be different from what we have estimated above.
However, assuming that tidally stripped stars wind up in a stellar
halo, we expect the total stellar mass lost to stripping to be small.
The Milky Way's thick disk contains a factor of $\sim10$ times more
stars than its stellar halo, and thus, any correction due to tidal
stripping is likely to be negligible.

While our mass-dependent blowout scenario explains our data well, it
is not clear that pre-galactic fragments actually suffer $\sim60\%$
baryon losses due to SN winds.  There are a wide range of results on
how effective SN winds should be at driving baryon outflow.  At the
one extreme, several groups argue that large SN driven outflows exist
in all galaxies with $V_c < 100$ km s$^{-1}$ \citep{Dekel86,Dekel03}.
At the the other extreme, simulations find that galaxies with $M >
10^6\msun$ experience almost no outflow \citep{MacLow99}.
Similarly, observational constraints on the extent of outflow vary.
\citet{Mayer04} use the baryonic Tully-Fisher relation to claim that
dwarf galaxies do not suffer large removal of baryons while
\citet{Strickland04} observe x-ray halos around massive star forming
galaxies ($M\sim 10^{10}-10^{11}\msun$) which suggest they
must have ejected at least some material.  Because we are considering
the role of blowout in low mass progenitors of our galaxy sample, we
claim that the current knowledge of gas blowout is moderately
consistent with our scenario and we await a more definitive
cosmological simulation which incorporates star formation and feedback
for detailed comparison to our model (Stinson et al., \emph{in
prep.}). 

There are several limitations with the simplified analysis we have
presented above.  First, we have ignored the difficult question of how
much material is accreted in continuous cold flows rather than bound
in halos \citep{Birnboim03,Keres04}.  Cold accretion of gas will tend
to increase the baryonic fraction of the thin disk.  Neglecting steady
gas accretion therefore leads us to overestimate the gas richness of
the merging pre-galactic fragments.  Second, we have not explicitly
considered how bulges are formed in the scenario discussed in
\S\ref{mergersec}, but we presume it involves repeated mergers of gas
rich sub-units with little net angular momentum, or a higher frequency
of major mergers in higher mass galaxies.  Within the sample we have
studies here, this omission is acceptable.  However, more theoretical
and observational work must be done to understand the thick disk
population in earlier type galaxies.

Finally, we find it difficult to reconcile the \citet{Brook04}
formation scenario with the increasing fraction of thick disk stars in
lower mass galaxies.  \citet{Brook04} suggest that thick disk stars
form {\it{in situ}} from high velocity dispersion gas during the
coalescence of sub-galactic fragments into a final disk.  However, we
see no obvious mechanism that could lead this scenario to produce a
larger fraction of thick disk stars in low mass galaxies.  One would
need to invoke a mechanism to increase the efficiency of star
formation at lower galaxy masses during mergers, while keeping star
formation inefficient at later times.  An alternative solution would
be if steady cold flow gas accretion is more important in massive
galaxies.  However, massive galaxies are more likely to have
established a hot shock-heated halo that would block cold flow
\citep[e.g.,][]{Dekel04}.  Thus, the likely behavior of cold flow
accretion has the opposite sign as what is needed to explain the high
baryonic fraction of thick disks in low mass galaxies.  Further
simulations will help constrain this and other possible solutions.

\subsubsection{The Scale Lengths of Thick \& Thin Disks}

Our data contribute to a growing number of observations finding that
thick disks have larger scale lengths than their embedded
thin disks \citep[see our Figure~\ref{2d_hrs}, and
Table~2]{Ojha01, Larsen03, Wu02, Pohlen04}.  The large
scale lengths of thick disks argue against their being formed via
vertical heating of a thin disk.  N-body simulations find that while
minor mergers can vertically heat a disk, they do not increase its
scale length \citep{Quinn93}.  Such mergers also tend to leave the
galaxy looking like an earlier Hubble type \citep{Walker96} while all
of our galaxies have no prominent bulge components.  As an example,
simulations by \citet{Aguerri01} find that minor mergers can extend
the scale length of the thin disk somewhat, by 10-60\%.  However, the
same simulations also produce a large bulge, which is incompatible
with our sample.

In contrast, in the accretion scenario one would expect the scale
length of the thin disk to be somewhat smaller than that of the thick
disk.  If the thin disk forms later from gas which has contracted
further into halo than the thick disk stars, it should have a smaller
scale length.  If angular momentum is largely conserved, then the thin
disk should also be rotating somewhat faster than the thick disk
because of its extra contraction.

The satellite accretion model therefore suggests that there may be
correlations between the radial scale lengths and the kinematics of
the thick and thin disks.  Results in Section~\ref{scale_length_sec}
and photometric decompositions by others \citep{Ojha01, Larsen03,
Wu02, Pohlen04} suggest that scale lengths of thick disks are roughly
30\% longer than those of their embedded thin disks, on average.
Simple angular momentum conservation would then suggest that the thick
disk should rotate with approximately 2/3 the speed of the thin disk,
in rough agreement with the Milky Way and FGC 1415 \citep{Yoachim05}.
However, the inclusion of any counter-rotating material in the merger
could easily break this correlation.  For example, the kinematics of
FGC 227 indicate that the satellites which contributed the majority of
the baryons to the thin disk could not also have deposited the
majority of the thick disk stars.  This particular formation pathway
allows the scale lengths of the thick and thin disks to sometimes
decouple, and indeed, the scale lengths of FGC 227's thick disk is
comparable to, not larger than, its thin disk.

The structural parameters of the thick disks formed in the
\citet{Abadi203} and Brook et al. (2004, 2005) simulations are in
moderate agreement with our results.  However, direct comparisons are
difficult because the simulated galaxies tend to be more massive than
the galaxies in our sample and also host large bulge components.  The
simulated thick disks do seem to match the observed trends of scale
height ratios (Figure~\ref{other_z}) and luminosity ratios
(Figure~\ref{other_l}).  However, the scale length ratio found in the
\citet{Brook05} simulation is fairly low (Figure~\ref{other_hr}),
possibly due to the fact that their thick disk stars are formed
directly from the gas during mergers, increasing the likelihood that
the thick and thin disk stars will share similar scale lengths and
kinematics.  It is also difficult to compare our 2-d decompositions
with analysis of simulations that can separate stellar populations
based on kinematics.

\subsection{Further Implications}

Given the excellent fit to the body of data on thick disks, we now
begin to address other implications of the accretion scenario
developed above.

\subsubsection{Old Low Mass Galaxies}

In hierarchical galaxy formation models, small scale structure
collapses first, suggesting that low mass galaxies should be old.
This expectation is in direct conflict with observations that low mass
galaxies almost always have blue colors consistent with young stellar
populations.  This difference is one of the most intractable failings
of the predictions of semi-analytic models \citep[e.g.][]{Bell03b,
vandenBosch02}.  The existence of thick disks that dominate the
stellar mass of low mass galaxies (Figure~\ref{massrat}) solves this
conundrum.  Our observations show that low mass galaxies are indeed
{\emph{dominated}} by an old stellar population, but one that is
sufficiently old, faint, and diffuse that it has no significant impact
on the observed colors of the young, high-surface brightness,
star-forming thin disk (\S\ref{csbs}).  We believe that semi-analytic
models could be brought into alignment with the data if they were to
include both the locking up of material into a diffuse thick disk and
the suppression of star formation efficiencies in low mass disks due
to their lying entirely below the Kennicutt star formation threshold
\citep[e.g.,][]{Verde02,Dalcanton04}.

\subsubsection{Abundance patterns and the timing of thick and thin disk
formation}

Studies of $\alpha$-element abundances of the Milky Way
have suggested that star formation in the thick disk took place over
several gigayears \citep[e.g.,][]{Bensby04b}.  The abundances in thick
disk stars show a flat plateau at high [$\alpha$/Fe] that extends to
[Fe/H]$\sim\!-0.3$, indicating that thick disk stars enriched quickly
to relatively high metallicity, before Type Ia supernovae became
prevalent.  At larger iron abundances ($-0.3<$[Fe/H]$\lesssim 0$),
however, the $\alpha$ abundance declines linearly, suggesting that
star formation in the thick disk was sufficiently extended ($\gtrsim
1-3\Gyr$) that enrichment from Type Ia's became important.  The
abundances of thin disk stars show similar, parallel behavior, but the
plateau does not extend to equally high metallicities, indicating that
early star formation in the thin disk was not nearly as rapid as in
the thick disk.  The abundance patterns of thick and thin disk stars
therefore follow parallel but distinct sequences on the [$\alpha$/Fe]
vs [Fe/H] plane, with significant overlap in [Fe/H] \citep[most
recently][]{Bensby04a,Mishenina04,Gratton03}.

The above sequence of events is typically taken as evidence that the
thick disk formed from violent heating of a previous thinner disk.
However, it may be possible to accommodate the abundance data in the
accretion scenario as well.  First, the rapid enrichment of future
thick disk stars can easily occur in the pre-galactic fragments.
These mini-halos should be dense, leading to high gas densities and
star formation rates, which would produce the necessary fast
enrichment.  While we have hypothesized above that supernova-driven
winds will truncate star formation in the lower mass progenitors, some
mini-halos will have sufficient mass to retain gas for longer periods
of time, allowing stars to form over sufficiently long timescales to
produce both the drop in [$\alpha$/Fe] and the enrichment of some
thick disk stars to near solar metallicities.  

The expected timescales for this scenario are compatible with the
observational constraints.  Theory suggests that the epoch of thick
disk assembly should correspond to the period of rapid mass accretion
seen in simulations at $z\!\gtrsim\! 3$, or $t_{lookback}\!\gtrsim\!
11\Gyr$ \citep[e.g.,][]{Zhao03}.  Observationally, the relative
abundance of [Eu/Ba] indicates that thick disk stars were formed on a
timescale of 1-1.5$\Gyr$ \citep{mashonkina03}, which makes the
theoretical expectation consistent with the age of the universe
determined from WMAP.  

In addition to the short star formation timescale for thick disk
stars, the accretion scenario can produce a long timescale for
formation of thin disk stars.  After the pre-galactic fragments merge
into a disk, the gas that forms the thin disk gradually converts into
stars.  The timescale of this conversion is controlled primarily by
the gas surface density.  In general, this timescale should be much
longer in the disk than in the pre-galactic fragments, because the gas
is spread over much larger areas, leading to lower gas densities and
longer star formation timescales.  The difference in timescales for
thick and thin disk star formation could lead to the appearance of a
``delay'' between the formation of the two components.  However, as
accretion of both gas and stars would be on-going from early times
until $z\sim3$, some genuinely old thick disk stars would be allowed
to form \citep[see discussion in][]{Abadi203}.

The accretion scenario also provides a mechanism for producing thin
disk stars with lower $\alpha$-abundances than thick disks stars at the
same metallicity.  Because the thin disk assembles from gas that had
not been consumed by star formation in pre-galactic fragments, it is
possible for the gas to initially have lower mean metallicity than the
thick disk stars that were accreted.  The gas may come from larger
radii within individual mini-halos, and thus be less enriched.  It may
also come from fragments that have never formed stars, or from cold
flow accretion directly.  Thus, accretion may allow the youngest thin
disk stars to be sufficiently metal-poor that they overlap the
metallicities of thick disk stars.

The one significant trouble spot is the thinness of the observed
[$\alpha$/Fe] vs [Fe/H] relation for thick disk stars.  If the thick
disk formed from assembly of many different sub-units of different
masses, lifetimes, and gas richnesses, then one might expect large
variations in the degree of $\alpha$-enhancement in the accreted
stars.  On the other hand, the potential discrepancy might not be as
severe as one might initially believe.  If supernova-driven winds
truncate star formation in low mass sub-units, then only the most
massive mini-halos contribute stars to the high metallicity
([Fe/H]$\gtrsim-0.3$) thick disk, since they are the only precursors
that could hold gas long enough to allow significant Type Ia
enrichment.  Massive halos are rarer than low mass halos, and thus a
relatively small number of halos may dominate the metal rich end of
the thick disk population, much in the way that $L_*$ galaxies
dominate the luminosity density of the local universe.  These
disrupted satellites may also segregate to different radii, as seen in
the \citet{Abadi203} simulations, such that a sample at the solar
circle is dominated by an even smaller number of massive satellites.
More detailed simulations are needed to evaluate the size of this
possible discrepancy.

\subsubsection{Pre-Enrichment of Thin Disks}

Chemical abundance data on stars within the Milky Way has led to the
conclusion that the thin disk may have been ``pre-enriched''
\citep[e.g.,][]{Caimmi00, Chiappini97, Pagel95}.  Such pre-enrichment
naturally explains the lack of truly metal poor stars in the thin disk
as well as the under-abundance of more moderately metal-poor stars
(i.e.\ the ``G-Dwarf'' problem).  In the scenario we have explored
here, the gas from which thin disk stars form was originally
associated with the thick disk, and thus will have been enriched while
still in pre-galactic fragments.  While this idea has been suggested
before \citep[e.g.,][]{Brook05}, the universality of thick disks
suggests that it is probably a wide-spread, phenomena.

\subsubsection{Producing the Mass-Metallicity Relationship in Disks}
%{\sc{Producing the Mass-Metallicity Relationship in Disks:}}
Another attractive feature of the satellite accretion model is that it
facilitates creating the mass-metallicity relationship in disks.  Many
authors have argued that the lower metallicities and effective yields
seen in low mass galaxies is due to the onset of supernova-driven
winds at the mass scale where the metallicity begins to fall
\citep[$V_c\!\sim\!120\kms$, or $M_{baryon}<3\times10^{10}\msun$;
e.g.][]{Garnett02,Tremonti04,Dekel03,Kauffmann03b}.  However,
simulations of gas outflow find that it is quite difficult to drive
coherent winds at these masses \citep{Ferrara00}, particularly given
the low star formation rates typical of low mass disks
\citep[e.g.][]{Hunter04}.

As an alternative, the satellite accretion model suggests that
non-negligible star formation took place in lower mass sub-units.
These pre-galactic fragments had much lower escape velocities, and
probably had higher gas surface densities due to not yet being
organized into a coherent rotating disk.  Thus, the sub-units are a
more natural environment for driving winds, given their low escape
velocities and likely high star formation rates.  The origin of the
observed mass-metallicity relation may therefore lie not so much in
the disks themselves, but in the sub-units from which they assembled.

%{\sc{Damped-Ly$\alpha$ Systems:}} Studies of damped Ly$\alpha$
%systems also support a thick disk formation during early
%merging/collapse stages of a galaxy.  \citet{Prochaska98} interpret
%large velocity widths in a sample of damped Ly$\alpha$ systems at high
%redshift as indicative of early structure formation in the form of
%thick rotating gas disks.  \citet{Wolfe98} find that if damped
%Ly$\alpha$ systems are thick rotating disks, they are metal rich
%enough to form stars with metallicities matching the MW thick disk
%while still leaving $\sim90\%$ of the baryons in gas which could then
%further contract to form a thin disk.

\section{Conclusions}

We fit thin and thick disk components to a sample of 34 late-type
edge-on spiral galaxies.  Our thick disk components are very similar
to previously detected thick disk systems and the MW thick disk,
suggesting they are a remnant stellar population left over from early
stages of galaxy formation.  In lower mass galaxies ($V_c < 100$ km
s$^{-1}$), the thick disk is the dominant component in both luminosity
and stellar mass.  For higher mass galaxies, the thick disk is a minor
component, and is analogous to the thick disks found in the Milky Way
and other higher mass galaxies. In particular, we find:
\begin{itemize}
\item{Thick disks have a scale height $\sim 2$ times larger than thin disks}
\item{Thick disks have systematically larger scale lengths than thin disks}
\item{In low mass galaxies, the thick disk can dominate the total
 $R$-band luminosity}
\item{The thick disk comprises 5-40\% of the total baryonic mass of
our galaxies}
\end{itemize}

We combine these results with the findings of other studies of thick
disks to analyze possible thick disk formation scenarios.  In
particular, we include results from thick disk kinematics
\citep{Yoachim05}, studies of resolved stellar populations in thick
disks \citep{Seth05,Mould05}, and simulations which form thick disks
\citep{Brook04,Brook05, Abadi203}.  Overall, we find that models where
the thick disk forms from a kinematically heated thin disk is not
supported by the data.  Instead, our results favor models where thick
disk stars formed in galactic sub-units before merging to create the
final galaxy.

We consider a hierarchical galaxy formation scenario where galaxies
form through a series of mergers where sub-units deposit both stars
and gas.  Any stellar component in the sub-units end up in the thick
disk, while gas cools and forms a thin disk.  We find that the low
mass galaxies in our sample must have formed from sub-units that had a
higher stellar mass fraction than those that formed higher mass
galaxies.  We can explain this result if low mass sub-units (which go
on to form low mass galaxies) are more susceptible to SN-induced
blowout, leaving them with a higher stellar to gas mass fraction.  A
mass-dependent blowout scenario is consistent with other general
observations of disk galaxies, such as the mass-metallicity relation
and the chemical pre-enrichment of the MW thin disk.

\begin{acknowledgments}
Thanks to Greg Stinson, Anil Seth, and Alyson Brooks for stimulating
conversations.  JJD and PY were partially supported through NSF grant
CAREER AST-0238683 and the Alfred P.\ Sloan Foundation.  This research
made extensive use of distributed computing with Condor software: The
Condor Software Program (Condor) was developed by the Condor Team at
the Computer Sciences Department of the University of
Wisconsin-Madison. All rights, title, and interest in Condor are owned
by the Condor Team.
\end{acknowledgments}

%\bibliography{/users/yoachim/Papers/Index/mybigbib}

%\include{tables/single_models}

%\include{tables/single_table}
\begin{deluxetable}{ c c c c c c c c c c c}
%\tabletypesize{\small \footnotesize \scriptsize}
\tabletypesize{\scriptsize}
%\rotate
%%%NEED TO PUT ROTATE BACK FOR NON-EMULATE VERSION
\tablewidth{0pt}
%\tablenum{num}
%\tablecolumns{19}
%\tableheadfrac{num}
\tablecaption{Single disk fits for the sample galaxies. \label{single_table}}
\tablehead{
 \colhead{} & \colhead{}& \multicolumn{3}{c}{\underline{~~~~~~~~~~~~~~~$B$~~~~~~~~~~~~~~~}} & 
\multicolumn{3}{c}{\underline{~~~~~~~~~~~~~~~$R$~~~~~~~~~~~~~~~}} &
\multicolumn{3}{c}{\underline{~~~~~~~~~~~~~~~$K_s$~~~~~~~~~~~~~~}} \\ 
%\cline{3-4} \cline{5-8} \cline{8-11}  \\
\colhead{FGC} & \colhead{Adopted Distance\tablenotemark{1}} & \colhead{$\mu(0,0)$} & \colhead{$h_r$} & \colhead{$z_0$} & \colhead{$\mu(0,0)$} & \colhead{$h_r$} & \colhead{$z_0$} &\colhead{$\mu(0,0)$} & \colhead{$h_r$} & \colhead{$z_0$} \\
 & Mpc  & (\surfb) & (\arcsec) & (\arcsec)  & (\surfb) & (\arcsec) & (\arcsec) & (\surfb) & (\arcsec) & (\arcsec)}
\startdata
31 & 51.9 & 22.67$^{-0.01}_{ 0.14}$ &   10.1$^{ 0.62}_{-0.61}$ &
  1.96$^{ 0.05}_{-0.10}$ & 21.93$^{-0.03}_{ 0.15}$ &    8.9$^{ 1.05}_{-0.20}$
&   2.06$^{ 0.05}_{-0.13}$ & 20.01$^{-0.01}_{ 0.19}$ &
   6.9$^{ 1.35}_{-0.00}$ &   1.95$^{ 0.21}_{-0.05}$  \\
36 & 80.9 & 22.33$^{-0.09}_{ 0.03}$ &    8.5$^{ 1.10}_{-0.37}$ &
  1.71$^{ 0.07}_{-0.04}$ & 21.06$^{-0.07}_{ 0.12}$ &    7.4$^{ 0.74}_{-0.41}$
&   1.73$^{ 0.12}_{-0.08}$ & 18.51$^{-0.02}_{ 0.15}$ &
   6.8$^{ 0.93}_{-0.06}$ &   1.64$^{ 0.08}_{-0.06}$  \\
130 & 233.1 & 22.67$^{-0.03}_{ 0.09}$ &    9.4$^{ 1.95}_{-0.12}$ &
  1.63$^{ 0.04}_{-0.00}$ & 21.15$^{-0.10}_{ 0.00}$ &    8.4$^{ 1.40}_{-0.07}$
&   1.71$^{ 0.07}_{-0.04}$ & 17.17$^{-0.01}_{ 0.16}$ &
   6.5$^{ 0.50}_{-0.19}$ &   1.27$^{ 0.05}_{-0.06}$  \\
164 & 69.9 & 22.88$^{-0.03}_{ 0.37}$ &   10.7$^{ 0.57}_{-0.28}$ &
  1.63$^{ 0.30}_{-0.12}$ & 22.24$^{-0.07}_{ 0.32}$ &    9.8$^{ 0.61}_{-0.17}$
&   1.86$^{ 0.32}_{-0.16}$ & 20.38$^{-0.02}_{ 0.01}$ &
   8.6$^{ 0.42}_{-0.31}$ &   1.87$^{ 0.02}_{-0.07}$  \\
215 & 131.1 & 22.46$^{-0.01}_{ 0.08}$ &   12.8$^{ 1.81}_{-0.33}$ &
  1.71$^{ 0.04}_{-0.02}$ & 21.21$^{-0.04}_{ 0.16}$ &   11.4$^{ 0.52}_{-0.63}$
&   1.64$^{ 0.11}_{-0.08}$ & 17.93$^{-0.10}_{ 0.12}$ &
   7.4$^{ 1.36}_{-0.91}$ &   1.23$^{ 0.01}_{-0.03}$  \\
225 & 74.3 & 22.29$^{-0.05}_{ 0.02}$ &    8.9$^{ 0.24}_{-0.72}$ &
  2.41$^{ 0.02}_{-0.07}$ & 21.31$^{-0.02}_{ 0.11}$ &    8.2$^{ 0.05}_{-0.61}$
&   2.40$^{ 0.07}_{-0.12}$ & 19.28$^{-0.02}_{ 0.18}$ &
   7.7$^{ 1.30}_{-0.03}$ &   2.60$^{ 0.17}_{-0.13}$  \\
227 & 89.4 & 22.52$^{-0.02}_{ 0.12}$ &   11.2$^{ 2.39}_{-0.21}$ &
  2.00$^{ 0.02}_{-0.01}$ & 21.21$^{-0.06}_{ 0.02}$ &   10.2$^{ 1.29}_{-0.10}$
&   2.05$^{ 0.05}_{-0.02}$ & 18.48$^{-0.07}_{ 0.20}$ &
   9.1$^{ 1.03}_{-0.42}$ &   2.01$^{ 0.16}_{-0.12}$  \\
277 & 84.9 & 23.14$^{-0.01}_{ 0.21}$ &    9.6$^{ 0.33}_{-0.14}$ &
  2.08$^{ 0.22}_{-0.09}$ & 21.75$^{-0.05}_{ 0.28}$ &    8.7$^{ 0.45}_{-0.06}$
&   2.24$^{ 0.30}_{-0.17}$ & 19.02$^{-0.08}_{ 0.28}$ &
   7.5$^{ 1.07}_{-0.38}$ &   2.02$^{ 0.31}_{-0.17}$  \\
310 & 80.8 & 22.79$^{-0.04}_{ 0.08}$ &    9.9$^{ 0.71}_{-0.23}$ &
  1.91$^{ 0.10}_{-0.04}$ & 21.19$^{-0.01}_{ 0.14}$ &    8.7$^{ 0.63}_{-0.29}$
&   1.95$^{ 0.11}_{-0.08}$ & 18.20$^{-0.10}_{ 0.28}$ &
   7.3$^{ 1.42}_{-0.62}$ &   1.70$^{ 0.17}_{-0.13}$  \\
349 & 117.6 & 22.21$^{-0.06}_{ 0.08}$ &    8.2$^{ 0.64}_{-0.46}$ &
  1.63$^{ 0.07}_{-0.04}$ & 21.09$^{-0.05}_{ 0.14}$ &    7.5$^{ 0.55}_{-0.34}$
&   1.71$^{ 0.11}_{-0.07}$ & 18.68$^{-0.02}_{ 0.21}$ &
   6.9$^{ 0.50}_{-0.06}$ &   1.75$^{ 0.18}_{-0.09}$  \\
395 & 109.3 & 22.95$^{-0.04}_{ 0.06}$ &   12.4$^{ 1.93}_{-0.09}$ &
  1.67$^{ 0.05}_{-0.02}$ & 21.46$^{-0.05}_{ 0.03}$ &   10.6$^{ 1.61}_{-0.02}$
&   1.74$^{ 0.03}_{-0.02}$ & 18.18$^{-0.11}_{ 0.12}$ &
   8.4$^{ 0.48}_{-1.06}$ &   1.42$^{ 0.07}_{-0.06}$  \\
436 & 109.2 & 22.58$^{-0.00}_{ 0.09}$ &    9.8$^{ 0.59}_{-0.00}$ &
  2.11$^{ 0.10}_{-0.02}$ & 21.06$^{-0.02}_{ 0.23}$ &    7.9$^{ 0.37}_{-0.16}$
&   2.11$^{ 0.20}_{-0.12}$ & 17.71$^{-0.14}_{ 0.25}$ &
   5.6$^{ 0.73}_{-0.60}$ &   1.55$^{ 0.14}_{-0.15}$  \\
446 & 88.2 & 22.43$^{-0.09}_{ 0.10}$ &   18.3$^{ 2.68}_{-0.57}$ &
  3.14$^{ 0.09}_{-0.09}$ & 20.74$^{-0.08}_{ 0.08}$ &   14.7$^{ 2.13}_{-0.08}$
&   3.03$^{ 0.08}_{-0.08}$ & 16.64$^{-0.15}_{ 0.39}$ &
  10.0$^{ 1.81}_{-0.72}$ &   1.89$^{ 0.27}_{-0.26}$  \\
780 & 34.4 & 22.22$^{-0.03}_{ 0.41}$ &   15.7$^{ 1.08}_{-0.94}$ &
  4.34$^{ 0.81}_{-0.61}$ & 21.41$^{-0.05}_{ 0.40}$ &   15.1$^{ 0.48}_{-0.95}$
&   4.96$^{ 0.88}_{-0.72}$ & 19.28$^{-0.00}_{ 0.02}$ &
  14.7$^{ 0.86}_{-0.01}$ &   3.95$^{ 0.01}_{-0.04}$  \\
901 & 131.2 & 22.30$^{-0.10}_{ 0.09}$ &    8.0$^{ 1.16}_{-0.48}$ &
  1.72$^{ 0.11}_{-0.07}$ & 21.10$^{-0.04}_{ 0.22}$ &    7.9$^{ 0.88}_{-0.71}$
&   1.71$^{ 0.15}_{-0.12}$ & 18.71$^{-0.03}_{ 0.05}$ &
   7.0$^{ 0.44}_{-0.29}$ &   1.52$^{ 0.02}_{-0.03}$  \\
913 & 62.5 & 21.98$^{-0.08}_{ 0.17}$ &    9.7$^{ 1.02}_{-0.60}$ &
  1.60$^{ 0.14}_{-0.09}$ & 21.04$^{-0.08}_{ 0.17}$ &    9.0$^{ 0.81}_{-0.60}$
&   1.73$^{ 0.14}_{-0.10}$ & 18.91$^{-0.05}_{ 0.03}$ &
   9.3$^{ 0.39}_{-0.03}$ &   1.86$^{ 0.02}_{-0.07}$  \\
979 & 52.0 & 21.35$^{-0.10}_{ 0.15}$ &   13.0$^{ 2.57}_{-0.18}$ &
  2.84$^{ 0.26}_{-0.24}$ & 20.27$^{-0.08}_{ 0.13}$ &   12.1$^{ 2.18}_{-0.27}$
&   3.03$^{ 0.23}_{-0.23}$ & 17.51$^{-0.06}_{ 0.29}$ &
  11.1$^{ 1.52}_{-0.01}$ &   2.53$^{ 0.32}_{-0.22}$  \\
1043 & 50.1 & 21.94$^{-0.05}_{ 0.08}$ &   20.7$^{ 3.01}_{-2.28}$ &
  3.38$^{ 0.20}_{-0.14}$ & 20.59$^{-0.03}_{ 0.14}$ &   16.9$^{ 0.38}_{-1.07}$
&   3.43$^{ 0.31}_{-0.17}$ & 16.91$^{-0.22}_{ 0.39}$ &
  10.6$^{ 1.22}_{-1.88}$ &   2.23$^{ 0.36}_{-0.36}$  \\
1063 & 56.4 & 22.08$^{-0.01}_{ 0.12}$ &    7.8$^{ 0.53}_{-0.23}$ &
  2.21$^{ 0.05}_{-0.10}$ & 21.19$^{-0.03}_{ 0.16}$ &    7.0$^{ 0.51}_{-0.17}$
&   2.22$^{ 0.10}_{-0.12}$ & 19.17$^{-0.02}_{ 0.07}$ &
   7.4$^{ 1.50}_{-0.02}$ &   2.18$^{ 0.01}_{-0.10}$  \\
1285 & 18.8 & 21.99$^{-0.06}_{ 0.25}$ &   22.6$^{ 1.20}_{-0.53}$ &
  6.05$^{ 0.55}_{-0.63}$ & 20.99$^{-0.11}_{ 0.26}$ &   19.7$^{ 1.74}_{-0.71}$
&   6.63$^{ 0.61}_{-0.75}$ & 18.59$^{-0.11}_{ 0.09}$ &
  15.8$^{ 0.10}_{-2.31}$ &   5.15$^{ 0.29}_{-0.23}$  \\
1303 & 51.7 & 22.57$^{-0.02}_{ 0.34}$ &    9.2$^{ 0.83}_{-0.35}$ &
  2.32$^{ 0.35}_{-0.24}$ & 21.70$^{-0.02}_{ 0.28}$ &    8.5$^{ 0.67}_{-0.40}$
&   2.50$^{ 0.30}_{-0.24}$ & 19.55$^{-0.02}_{ 0.11}$ &
   5.8$^{ 1.17}_{-0.06}$ &   2.56$^{ 0.18}_{-0.10}$  \\
1415 & 38.3 & 21.79$^{-0.04}_{ 0.31}$ &   19.1$^{ 1.88}_{-0.73}$ &
  3.84$^{ 0.56}_{-0.40}$ & 20.83$^{-0.05}_{ 0.37}$ &   18.3$^{ 0.39}_{-1.18}$
&   4.27$^{ 0.69}_{-0.54}$ & 18.34$^{-0.01}_{ 0.18}$ &
  15.3$^{ 1.43}_{-0.00}$ &   3.21$^{ 0.29}_{-0.15}$  \\
1440 & 70.9 & 22.04$^{-0.07}_{ 0.02}$ &   19.7$^{ 2.05}_{-1.36}$ &
  2.74$^{ 0.15}_{-0.10}$ & 20.54$^{-0.05}_{ 0.20}$ &   15.9$^{ 0.86}_{-0.15}$
&   2.78$^{ 0.22}_{-0.23}$ & 16.81$^{-0.12}_{ 0.26}$ &
  10.2$^{ 1.61}_{-1.00}$ &   1.83$^{ 0.15}_{-0.15}$  \\
1642 & 36.6 & 22.60$^{-0.04}_{ 0.10}$ &   12.2$^{ 1.16}_{-0.85}$ &
  3.04$^{ 0.12}_{-0.13}$ & 21.76$^{-0.01}_{ 0.24}$ &   12.5$^{ 1.32}_{-0.14}$
&   3.53$^{ 0.33}_{-0.26}$ & 19.94$^{-0.07}_{ 0.04}$ &
  18.5$^{ 5.02}_{-5.11}$ &   3.14$^{ 0.05}_{-0.08}$  \\
1948 & 36.9 & 22.67$^{-0.03}_{ 0.27}$ &   13.1$^{ 0.40}_{-0.87}$ &
  2.70$^{ 0.27}_{-0.26}$ & 21.86$^{-0.04}_{ 0.22}$ &   12.3$^{ 0.51}_{-0.44}$
&   2.98$^{ 0.24}_{-0.25}$ & 19.76$^{-0.10}_{ 0.00}$ &
   8.7$^{ 0.06}_{-2.92}$ &   2.31$^{ 0.01}_{-0.00}$  \\
2131 & 41.7 & 22.51$^{-0.08}_{ 0.08}$ &   10.7$^{ 1.53}_{-0.33}$ &
  3.15$^{ 0.18}_{-0.13}$ & 21.30$^{-0.05}_{ 0.10}$ &   10.0$^{ 1.06}_{-0.32}$
&   3.46$^{ 0.18}_{-0.16}$ & 18.62$^{-0.00}_{ 0.05}$ &
   8.9$^{ 0.28}_{-0.03}$ &   3.06$^{ 0.10}_{-0.03}$  \\
2135 & 125.3 & 22.31$^{-0.04}_{ 0.06}$ &    7.6$^{ 1.00}_{-0.29}$ &
  1.67$^{ 0.00}_{-0.02}$ & 21.06$^{-0.06}_{ 0.15}$ &    6.9$^{ 0.67}_{-0.23}$
&   1.73$^{ 0.08}_{-0.09}$ & 18.08$^{-0.05}_{ 0.14}$ &
   4.8$^{ 0.28}_{-0.20}$ &   1.51$^{ 0.09}_{-0.06}$  \\
2369 & 59.8 & 22.75$^{-0.07}_{ 0.17}$ &    8.8$^{ 1.11}_{-0.34}$ &
  1.90$^{ 0.21}_{-0.12}$ & 21.81$^{-0.03}_{ 0.30}$ &    8.7$^{ 0.62}_{-0.50}$
&   2.14$^{ 0.32}_{-0.16}$ & 19.80$^{-0.02}_{ 0.06}$ &
   9.5$^{ 1.27}_{-0.21}$ &   2.09$^{ 0.04}_{-0.04}$  \\
2548 & 55.6 & 22.75$^{-0.05}_{ 0.23}$ &   10.7$^{ 1.54}_{-0.03}$ &
  2.17$^{ 0.30}_{-0.15}$ & 21.65$^{-0.02}_{ 0.24}$ &    9.9$^{ 0.87}_{-0.05}$
&   2.43$^{ 0.29}_{-0.19}$ & 19.39$^{-0.00}_{ 0.19}$ &
   9.6$^{ 0.76}_{-0.01}$ &   2.51$^{ 0.27}_{-0.09}$  \\
2558 & 73.8 & 22.29$^{-0.03}_{ 0.07}$ &    9.8$^{ 1.27}_{-0.64}$ &
  3.06$^{ 0.03}_{-0.14}$ & 21.27$^{-0.02}_{ 0.18}$ &    9.2$^{ 1.00}_{-0.44}$
&   3.15$^{ 0.14}_{-0.24}$ & 19.05$^{-0.05}_{ 0.13}$ &
   9.1$^{ 1.41}_{-0.45}$ &   2.82$^{ 0.13}_{-0.12}$  \\
E1371 & 82.6 & 23.02$^{-0.10}_{ 0.14}$ &    8.7$^{ 1.51}_{-1.07}$ &
  2.12$^{ 0.08}_{-0.02}$ & 21.12$^{-0.02}_{ 0.10}$ &    7.7$^{ 1.04}_{-0.44}$
&   2.07$^{ 0.03}_{-0.04}$ & 17.06$^{-0.02}_{ 0.19}$ &
   6.8$^{ 0.19}_{-0.25}$ &   1.47$^{ 0.12}_{-0.09}$  \\
E1404 & 76.2 & 22.60$^{-0.04}_{ 0.06}$ &    8.9$^{ 0.67}_{-0.53}$ &
  1.58$^{ 0.03}_{-0.03}$ & 21.36$^{-0.05}_{ 0.23}$ &    7.8$^{ 0.68}_{-0.12}$
&   1.64$^{ 0.12}_{-0.10}$ & 18.76$^{-0.12}_{ 0.18}$ &
   7.0$^{ 0.54}_{-0.84}$ &   1.64$^{ 0.13}_{-0.11}$  \\
E1498 & 135.5 & 22.48$^{-0.10}_{ 0.04}$ &    8.3$^{ 1.89}_{-0.21}$ &
  1.48$^{ 0.10}_{-0.02}$ & 21.03$^{-0.12}_{ 0.01}$ &    7.6$^{ 1.51}_{-0.05}$
&   1.51$^{ 0.09}_{-0.04}$ & 17.50$^{-0.01}_{ 0.32}$ &
   6.7$^{ 0.38}_{-0.35}$ &   1.11$^{ 0.14}_{-0.07}$  \\
E1623 & 261.1 & 22.70$^{-0.04}_{ 0.08}$ &    7.9$^{ 0.85}_{-0.08}$ &
  1.35$^{ 0.03}_{-0.02}$ & 21.07$^{-0.01}_{ 0.10}$ &    6.4$^{ 0.18}_{-0.07}$
&   1.25$^{ 0.05}_{-0.04}$ & 17.37$^{-0.08}_{ 0.17}$ &
   4.5$^{ 0.28}_{-0.20}$ &   0.96$^{ 0.06}_{-0.07}$  \\
\enddata
\tablenotetext{1}{
These fits use Equations~\ref{z3} and~\ref{sech2} with $N=1$ (i.e. a
sech$^2$ vertical profile).  Peak edge-on surface brightnesses have not been
corrected for inclination.  When available, distances  taken from
\citet{Kara00}.  Otherwise, we have used the recessional velocity
corrected for Local Group infall to the Virgo cluster
(LEDA). Throughout, we assume $H_0=70$ km s$^{-1}$ Mpc$^{-1}$.
}

\end{deluxetable}

\begin{deluxetable}{  l l l}
%\tabletypesize{\small \footnotesize \scriptsize}
\tabletypesize{\small}
%\rotate
\tablewidth{5in}
%\tablenum{num}
%\tablecolumns{19}
%\tableheadfrac{num}
\tablecaption{Vertical light profiles for the two disk models used in fitting $R$-band structural parameters. \label{vert_doub}}
\tablehead{
 \colhead{Thin disk model} & \colhead{Thick disk model} & {Notes} 
}
\startdata
 $\mathrm{sech}^{2}(z/z_{0})$ & $\mathrm{sech}^{2}(z/z_{0})$ &  convolved w/1\arcsec ~FWHM Gaussian  \\
 $\mathrm{sech}^{2}(z/z_{0})$ & $\mathrm{sech}^{2}(z/z_{0})$ & midplane masked\\
 $\mathrm{sech}^{2}(z/z_{0})$ & $\mathrm{sech}^{2}(z/z_{0})$ &  \\
 $\mathrm{sech}(z/z_{0})$ & $\mathrm{sech}^{2}(z/z_{0})$ &  \\
 $\mathrm{sech}^{2}(z/z_{0})$ & $\mathrm{sech}(z/z_{0})$ &  \\
 $\mathrm{sech}(z/z_{0})$ & $\mathrm{sech}(z/z_{0})$ &  \\
\enddata
\end{deluxetable}

\begin{deluxetable}{ c c c c c c c c c}
%\tabletypesize{\small \footnotesize \scriptsize}
\tabletypesize{\scriptsize}
%\rotate
\tablewidth{6.25in}
%\tablenum{num}
%\tablecolumns{num}
%\tableheadfrac{num}
%\tablecaption{text \label{key}}
\tablehead{
 & \multicolumn{3}{c}{ \underline{~~~~~~~~~~~~Thin Disk~~~~~~~~~~~~}} & \multicolumn{3}{c}{\underline{~~~~~~~~~~~~Thick Disk~~~~~~~~~~    }} \\
\colhead{FGC} & \colhead{$\mu(0,0)$} &  \colhead{$h_r$} &  \colhead{$z_0$} &  \colhead{$\mu(0,0)$} &   \colhead{$h_r$} &  \colhead{$z_0$} & \colhead{$L_{thick}/L_{thin}$ } & \colhead{n converged}\\
 & (mag/$\square$\arcsec)  & (\arcsec) & (\arcsec) &    (mag/$\square$\arcsec)  & (\arcsec) & (\arcsec)   
}
\tablecaption{
Two disk fits to $R$-band images.  Fit parameters are 
the best values found for each disk having a sech$^2$ vertical profile.  
The listed values are medians with uncertainties indicating the full range of convergent 
fits.  \label{two_disk_table}}
\startdata
31 & 22.0$^{+0.40}_{-0.03}$ &    8.0$^{+0.7}_{-4.2}$ &    1.7$^{+0.0}_{-0.9}$
& 23.8$^{+0.57}_{-1.60}$ &   11.8$^{+0.1}_{-0.6}$ &    2.9$^{+0.6}_{-0.8}$ &
 0.40$^{+0.00}_{-0.16}$ & 3  \\
36 & 22.2$^{+0.53}_{-0.21}$ &    6.8$^{+0.1}_{-0.5}$ &    1.0$^{+0.5}_{-0.3}$
& 21.4$^{+0.13}_{-0.13}$ &    7.4$^{+0.0}_{-0.1}$ &    1.9$^{+0.0}_{-0.1}$ &
 4.20$^{+1.71}_{-3.46}$ & 4  \\
130 & 21.1$^{+0.92}_{-0.03}$ &    8.6$^{+0.4}_{-0.2}$ &    1.6$^{+0.4}_{-0.2}$
& 24.1$^{+0.68}_{-1.31}$ &    9.5$^{+1.1}_{-1.0}$ &    3.9$^{+1.1}_{-1.8}$ &
 0.25$^{+0.08}_{-0.14}$ & 5  \\
164 & 22.4$^{+0.63}_{-0.04}$ &    9.5$^{+0.1}_{-0.5}$ &    1.5$^{+0.4}_{-0.4}$
& 23.7$^{+1.00}_{-0.17}$ &   12.2$^{+0.2}_{-1.4}$ &    4.7$^{+1.0}_{-1.3}$ &
 0.72$^{+0.30}_{-0.44}$ & 5  \\
215 & 21.3$^{+0.16}_{ 0.00}$ &   10.3$^{+0.0}_{-1.3}$ &    1.4$^{+0.0}_{-0.3}$
& 22.1$^{+0.83}_{-0.00}$ &   12.1$^{+0.0}_{-0.9}$ &    2.8$^{+0.0}_{-0.7}$ &
 0.23$^{+0.23}_{-0.15}$ & 3  \\
225 & 21.3$^{+2.43}_{ 0.00}$ &    7.6$^{+0.0}_{-2.3}$ &    2.1$^{+0.0}_{-1.4}$
& 21.3$^{+2.63}_{ 0.00}$ &    8.5$^{+0.0}_{-0.8}$ &    3.8$^{+0.0}_{-1.3}$ &
 0.39$^{+0.00}_{-0.24}$ & 3  \\
227 & 21.3$^{+0.91}_{-0.12}$ &   10.8$^{+1.0}_{-0.8}$ &    1.8$^{+0.5}_{-0.2}$
& 22.7$^{+1.07}_{-0.74}$ &   10.1$^{+2.2}_{-0.8}$ &    3.9$^{+0.1}_{-1.4}$ &
 0.26$^{+0.12}_{-0.19}$ & 5  \\
277 & 21.9$^{+0.59}_{-0.12}$ &    8.0$^{+0.3}_{-0.1}$ &    1.7$^{+0.7}_{-0.1}$
& 23.5$^{+0.95}_{-0.34}$ &   11.3$^{+1.6}_{-0.8}$ &    4.5$^{+1.0}_{-0.8}$ &
 0.47$^{+0.52}_{-0.24}$ & 5  \\
310 & 21.3$^{+0.17}_{-0.00}$ &    8.4$^{+0.0}_{-0.6}$ &    1.6$^{+0.0}_{-0.3}$
& 22.5$^{+0.21}_{ 0.00}$ &    9.6$^{+0.0}_{-0.5}$ &    2.9$^{+0.0}_{-0.0}$ &
 0.55$^{+0.17}_{-0.14}$ & 3  \\
349 & 21.2$^{+0.85}_{-0.01}$ &    7.0$^{+0.2}_{-0.4}$ &    1.4$^{+0.4}_{-0.5}$
& 22.3$^{+0.46}_{-0.80}$ &    7.3$^{+0.0}_{-0.1}$ &    2.4$^{+0.2}_{-0.5}$ &
 0.62$^{+2.05}_{-0.20}$ & 5  \\
395 & 21.3$^{+0.20}_{ 0.00}$ &   11.0$^{+0.0}_{-0.7}$ &    1.6$^{+0.0}_{-0.3}$
& 24.8$^{+0.38}_{ 0.00}$ &   11.2$^{+0.0}_{-0.4}$ &    6.4$^{+0.0}_{-0.2}$ &
 0.07$^{+0.03}_{-0.02}$ & 3  \\
436 & 21.1$^{+0.66}_{-0.04}$ &    7.3$^{+0.3}_{-0.1}$ &    1.7$^{+0.7}_{-0.1}$
& 23.0$^{+1.62}_{-0.15}$ &    9.9$^{+2.9}_{-0.1}$ &    4.2$^{+1.9}_{-0.4}$ &
 0.40$^{+0.10}_{-0.29}$ & 5  \\
446 & 20.8$^{+0.08}_{-0.16}$ &   14.5$^{+0.5}_{-0.0}$ &    2.9$^{+0.0}_{-0.2}$
& 23.9$^{+0.52}_{-0.49}$ &   16.2$^{+5.4}_{-0.6}$ &    4.6$^{+2.9}_{-0.2}$ &
 0.14$^{+0.43}_{-0.05}$ & 4  \\
780 & 21.6$^{+0.63}_{-0.07}$ &   13.4$^{+0.7}_{-0.4}$ &    3.1$^{+1.3}_{-0.4}$
& 22.6$^{+0.58}_{-0.13}$ &   16.1$^{+1.3}_{-0.3}$ &    8.4$^{+0.4}_{-1.1}$ &
 0.93$^{+0.57}_{-0.34}$ & 5  \\
901 & 21.2$^{+0.07}_{-0.14}$ &    6.9$^{+0.1}_{-0.1}$ &    1.3$^{+0.1}_{-0.2}$
& 23.1$^{+0.24}_{-0.96}$ &    8.8$^{+0.3}_{-0.1}$ &    2.9$^{+0.3}_{-0.6}$ &
 0.39$^{+0.02}_{-0.06}$ & 4  \\
913 & 21.2$^{+0.65}_{-0.22}$ &    7.8$^{+1.0}_{-0.5}$ &    1.4$^{+0.4}_{-0.1}$
& 22.4$^{+0.54}_{-0.01}$ &    9.1$^{+0.8}_{-0.5}$ &    2.5$^{+0.4}_{-0.1}$ &
 0.52$^{+0.13}_{-0.14}$ & 5  \\
979 & 20.2$^{+0.76}_{-0.08}$ &   11.9$^{+0.2}_{-0.5}$ &    2.3$^{+1.0}_{-0.8}$
& 21.7$^{+1.21}_{-0.69}$ &   13.0$^{+0.7}_{-0.6}$ &    5.1$^{+0.2}_{-1.2}$ &
 0.67$^{+0.82}_{-0.49}$ & 5  \\
1043 & 20.8$^{+0.05}_{-0.11}$ &   19.3$^{+0.0}_{-1.1}$ &
   2.7$^{+0.0}_{-0.1}$ & 22.3$^{+0.57}_{-0.17}$ &   11.3$^{+1.7}_{-0.5}$ &
   6.9$^{+0.5}_{-0.8}$ &  0.33$^{+0.06}_{-0.21}$ & 5  \\
1063 & 22.3$^{+0.04}_{-1.32}$ &    4.8$^{+0.1}_{-0.1}$ &
   0.8$^{+0.4}_{-0.1}$ & 21.5$^{+0.01}_{-0.12}$ &    7.3$^{+0.4}_{-0.2}$ &
   2.4$^{+0.0}_{-0.1}$ &  6.91$^{+4.69}_{-1.65}$ & 4  \\
1285 & 21.2$^{+0.29}_{-0.05}$ &   17.3$^{+1.6}_{-1.2}$ &
   4.4$^{+0.7}_{-0.3}$ & 22.2$^{+0.58}_{-0.02}$ &   23.6$^{+1.8}_{-1.0}$ &
  10.1$^{+0.6}_{-0.7}$ &  1.08$^{+1.00}_{-0.45}$ & 5  \\
1303 & 22.2$^{+0.52}_{-0.06}$ &    7.9$^{+0.0}_{-1.4}$ &
   1.4$^{+0.7}_{-0.3}$ & 22.4$^{+0.53}_{-0.13}$ &    9.0$^{+0.3}_{-0.3}$ &
   3.3$^{+0.3}_{-0.3}$ &  1.93$^{+1.82}_{-1.03}$ & 5  \\
1415 & 20.9$^{+0.63}_{-0.30}$ &   15.0$^{+0.6}_{-1.6}$ &
   2.8$^{+1.0}_{-0.5}$ & 22.1$^{+0.68}_{-0.07}$ &   21.1$^{+1.6}_{-1.3}$ &
   6.6$^{+0.8}_{-0.4}$ &  0.95$^{+0.36}_{-0.43}$ & 5  \\
1440 & 20.6$^{+0.05}_{-0.02}$ &   15.7$^{+0.4}_{-0.3}$ &
   2.3$^{+0.1}_{-0.1}$ & 22.7$^{+0.11}_{-0.16}$ &   17.2$^{+0.1}_{-0.2}$ &
   5.0$^{+0.1}_{-0.2}$ &  0.38$^{+0.17}_{-0.05}$ & 4  \\
1642 & 21.8$^{+0.58}_{-0.16}$ &   11.6$^{+0.4}_{-0.4}$ &
   3.1$^{+0.8}_{-0.4}$ & 24.6$^{+0.65}_{-1.05}$ &   19.5$^{+0.9}_{-4.7}$ &
  10.0$^{+1.3}_{-4.0}$ &  0.19$^{+0.33}_{-0.08}$ & 5  \\
1948 & 22.5$^{+0.69}_{-0.50}$ &   10.1$^{+0.8}_{-0.4}$ &
   1.6$^{+0.5}_{-0.3}$ & 22.4$^{+0.07}_{-0.06}$ &   13.0$^{+0.5}_{-0.3}$ &
   3.6$^{+0.0}_{-0.1}$ &  3.56$^{+2.22}_{-1.00}$ & 4  \\
2131 & 21.3$^{+0.22}_{-0.00}$ &    9.3$^{+0.2}_{-0.7}$ &
   2.8$^{+0.1}_{-0.8}$ & 22.7$^{+0.88}_{-0.64}$ &   10.5$^{+1.3}_{-0.1}$ &
   4.9$^{+1.3}_{-0.5}$ &  0.29$^{+0.33}_{-0.00}$ & 4  \\
2135 & 21.2$^{+0.21}_{-0.09}$ &    6.5$^{+0.7}_{-0.5}$ &
   1.1$^{+0.4}_{-0.0}$ & 22.3$^{+1.62}_{-0.29}$ &    8.5$^{+0.1}_{-1.0}$ &
   2.5$^{+1.3}_{-0.1}$ &  0.88$^{+0.79}_{-0.67}$ & 4  \\
2369 & 22.2$^{+0.52}_{-0.40}$ &    8.3$^{+0.1}_{-0.5}$ &
   1.4$^{+0.6}_{-0.0}$ & 23.3$^{+0.14}_{-0.44}$ &    9.3$^{+0.2}_{-1.0}$ &
   3.4$^{+0.2}_{-0.4}$ &  0.75$^{+0.46}_{-0.21}$ & 3  \\
2548 & 21.9$^{+0.05}_{-0.08}$ &    9.9$^{+0.1}_{-0.1}$ &
   1.4$^{+0.2}_{-0.1}$ & 22.6$^{+0.09}_{-0.12}$ &    9.8$^{+0.0}_{-0.0}$ &
   3.5$^{+0.1}_{-0.1}$ &  1.18$^{+0.76}_{-0.14}$ & 4  \\
2558 & 21.8$^{+0.06}_{-0.56}$ &    8.4$^{+1.3}_{-0.7}$ &
   2.6$^{+0.1}_{-0.4}$ & 22.4$^{+0.93}_{-0.41}$ &   10.1$^{+0.0}_{-0.2}$ &
   3.6$^{+1.2}_{-0.0}$ &  0.47$^{+1.31}_{-0.17}$ & 4  \\
E1371 & 21.2$^{+0.91}_{-0.12}$ &    8.6$^{+0.3}_{-1.2}$ &
   1.6$^{+0.7}_{-0.0}$ & 22.9$^{+0.17}_{-0.49}$ &    7.3$^{+0.1}_{-1.4}$ &
   3.4$^{+0.0}_{-0.7}$ &  0.27$^{+0.37}_{-0.04}$ & 5  \\
E1404 & 21.6$^{+1.09}_{-0.06}$ &    6.8$^{+0.2}_{-0.3}$ &
   1.3$^{+0.1}_{-0.2}$ & 22.4$^{+0.26}_{-0.38}$ &    9.2$^{+0.4}_{-0.3}$ &
   2.2$^{+0.1}_{-0.2}$ &  1.12$^{+2.04}_{-0.53}$ & 4  \\
E1498 & 20.9$^{+0.11}_{-0.05}$ &    7.7$^{+0.3}_{-0.2}$ &
   1.2$^{+0.2}_{-0.1}$ & 23.8$^{+0.35}_{-0.91}$ &    8.3$^{+0.2}_{-0.2}$ &
   3.8$^{+0.4}_{-1.0}$ &  0.19$^{+0.05}_{-0.04}$ & 4  \\
E1623 & 21.3$^{+0.11}_{-0.15}$ &    6.4$^{+0.1}_{-0.5}$ &
   0.9$^{+0.2}_{-0.0}$ & 22.5$^{+0.23}_{-0.53}$ &    6.3$^{+0.1}_{-0.2}$ &
   1.8$^{+0.0}_{-0.2}$ &  0.53$^{+0.20}_{-0.12}$ & 4  
\enddata
\end{deluxetable}

\end{document}